\documentclass[onecolumn, 12pt]{IEEEtran}

\usepackage{amsmath}
\usepackage{amssymb}
\usepackage[dvips]{graphicx}
\usepackage{setspace}
\usepackage{epsfig}
\newcommand{\K}{{\mathbf{K}}}
\newcommand{\x}{\mathbf{x}}
\newcommand{\y}{\mathbf{y}}

\newcommand{\n}{\mathbf{n}}
\newcommand{\hh}{\mathbf{H}}

\newcommand{\I}{\mathbf{I}}

\newcommand{\E}{\mathbb{E}}

\newcommand{\R}{{\sf{R}}}

\newcommand{\C}{{\sf{C}}}

\newcommand{\dd}{\dagger}

\newcommand{\tr}{{\text{tr\,}}}
\newcommand{\0}{\mathbf{0}}
\newcommand{\bu}{\mathbf{u}}
\newcommand{\U}{\mathbf{U}}
\newcommand{\bLambda}{\mathbf{\Lambda}}
\newcommand{\bPhi}{\mathbf{\Phi}}
\newcommand{\lm}{\lambda_{\max}}

\newcommand{\so}{\mathcal{S}_0}
\newcommand{\id}{{\text{id}}}
\newcommand{\G}{\mathbf{G}}
\newcommand{\si}{\mathcal{S}_\infty}
\newcommand{\li}{\mathcal{L}_\infty}
\newcommand{\Ri}{R_{E, \text{id}}}

\newcommand{\figsize}{0.7}

\newcommand{\tsnr}{{\text{\footnotesize{SNR}}}}

\newcommand{\ssnr}{\text{\scriptsize{SNR}}}
\newcommand{\tisnr}{\text{\tiny{SNR}}}

\newtheorem{theo}{Theorem}

\newtheorem{rem}{Remark}
\newtheorem{cor}{Corollary}

\begin{document}

\title{MIMO Wireless Communications under Statistical Queueing Constraints}



%
\author{\vspace{1cm}
\authorblockN{Mustafa Cenk Gursoy}
\thanks{The author is with the Department of Electrical
Engineering, University of Nebraska-Lincoln, Lincoln, NE, 68588
(e-mail: gursoy@engr.unl.edu).}
\thanks{This work was supported by the National Science Foundation under Grants CCF -- 0546384 (CAREER), CNS -- 0834753, and CCF-0917265. The material in this paper will be  presented in part at the Forty-Seventh Annual Allerton Conference on Communication, Control, and Computing in Oct. 2009.
}}


\maketitle
\thispagestyle{empty}

\begin{spacing}{1.5}

\begin{abstract}
The performance of multiple-input multiple-output wireless systems is investigated in the presence of statistical queueing constraints. Queuing constraints are imposed as limitations on buffer violation probabilities. The performance under such constraints is captured through the effective capacity formulation. A detailed analysis of the effective capacity is carried out in the low-power, wideband, and high--signal-to-noise ratio ($\tsnr$) regimes. In the low-power analysis, expressions for the first and second derivatives of the effective capacity with respect to $\tsnr$ at $\tsnr = 0$ are obtained under various assumptions on the degree of channel state information at the transmitter. Transmission strategies that are optimal in the sense of achieving the first and second derivatives are identified. It is shown that while the first derivative does not get affected by the presence of queueing constraints, the second derivative gets smaller as the constraints become more stringent. Through the energy efficiency analysis, this is shown to imply that the minimum bit energy requirements do not change with more strict limitations but the wideband slope diminishes. Similar results are obtained in the wideband regime if rich multipath fading is being experienced. On the other hand, sparse multipath fading with bounded number of degrees of freedom is shown to increase the minimum bit energy requirements in the presence of queueing constraints. Following the low-$\tsnr$ study, the impact of buffer limitations on the high-$\tsnr$ performance is quantified by analyzing the high-$\tsnr$ slope and the power offset in Rayleigh fading channels. Finally, numerical results are provided to illustrate the theoretical findings, and to demonstrate the interactions between the queueing constraints and spatial dimensions over a wide range of $\tsnr$ values.
\end{abstract}

\section{Introduction}

Having multiple antennas at the transmitter and receiver has been shown to improve the performance significantly in terms of both reliability and throughput when the channel fading coefficients are known at the receiver and/or transmitter. Due to these promising gains in the performance, information-theoretic analysis of multiple-input multiple-output (MIMO) channels has attracted much interest in the research community. In particular, considerable effort has been expended in the study of the capacity of MIMO wireless channels (see e.g., \cite{goldsmith} and the references therein). For instance, multiple-antenna capacity is studied in the low-power regime in \cite{verdu} and \cite{lozano}, and in the high-$\tsnr$ regime in \cite{lozano-highsnr}. In most studies on MIMO channel capacity, ergodic Shannon capacity formulation is employed as the main performance metric. However, this formulation does not capture the performance in the presence of quality-of-service (QoS) limitations in the form of constraints on queueing delays or queue lengths, although providing QoS assurances is of paramount importance in many delay-sensitive wireless systems, e.g., voice over IP (VoIP), and interactive and streaming video applications.

In \cite{dapeng}, effective capacity is proposed  as a metric that can be employed to measure the performance in the presence of statistical QoS limitations. Effective
capacity formulation uses the large deviations theory and
incorporates the statistical QoS constraints by capturing the rate
of decay of the buffer occupancy probability for large queue
lengths. Hence, effective capacity can be regarded as the maximum
throughput of a system operating under limitations on the buffer
violation probability. This formulation is tightly linked and in a sense dual to the concept of effective bandwidth \cite{chang} \cite{cschang} that is employed in the analysis of how much resource in terms of service rates is needed to support a given time-varying arrival process. The analysis of the effective capacity in various wireless communication settings has been conducted in several recent studies (see e.g., \cite{tang-powerrate} -- \cite{imperfect}).

In this paper, we study the effective capacity of MIMO wireless channels. In particular, we consider the low-power, wideband, and high-$\tsnr$ regimes and identify the impact of the QoS limitations\footnote{Throughout the paper, we use the terms ``QoS constraints", ``queueing constraints", and ``buffer constraints" interchangeably.} on the performance. We would like to note that recently references  \cite{Jorswieck}  and \cite{liu-isit} have also investigated the effective capacity of multiple-antenna channels. In \cite{Jorswieck}, the authors study the multiple-input single-output (MISO) channels and determine the optimal transmit strategies with covariance feedback. In \cite{liu-isit}, the concentration is on the MISO and single-input multiple-output (SIMO) channels. Analysis of MIMO channels is carried out only in the large antenna regime in which the number of receive and/or transmit antennas increase without bound. In addition, the authors in \cite{liu-isit} consider a MIMO channel matrix with independent and identically distributed (i.i.d.) zero-mean Gaussian entries, and consider equal power allocation across the antennas. In this paper, we consider a general MIMO link model in which the fading coefficients have arbitrary distributions and are possibly correlated\footnote{Only in the high-$\tsnr$ regime, we concentrate on the canonical MIMO model in which the fading coefficients are i.i.d. zero-mean, unit-variance, Gaussian random variables.}, provide a detailed study of the low-power, wideband, and high-$\tsnr$ regimes, investigate the transmission strategies under various assumptions on the degree of channel knowledge at the transmitter, and identify the impact of QoS constraints on the performance. The original contributions of this paper are the following:
\begin{enumerate}
\item We obtain expressions for the first and second derivatives of the effective capacity at $\tsnr = 0$ under various assumptions on the availability of channel knowledge at the transmitter, and show that while the first derivative is independent of the queueing constraints, the second derivative diminishes as the constraints become more stringent. Transmission strategies that achieve these derivatives are identified.

\item As a result of the findings on the derivatives of the effective capacity, we determine in the low-power regime that the minimum bit energy requirements in the presence of QoS limitations are the same as those attained in the absence of such constraints. On the other hand, we show that the wideband slope decreases under more strict queueing constraints, indicating that energy expenditure increases unless one is operating at the minimum bit energy level.

\item Under certain assumptions, we show that the results obtained in the low-power regime apply to the wideband regime with rich multipath fading. In contrast, we establish that sparse multipath fading has a significant impact on the performance in the wideband regime. In particular, we prove that minimum bit energies greater than that achieved in the absence of QoS constraints are required if the number of degrees of freedom in the form of noninteracting subchannels remain bounded as the bandwidth increases.

\item Considering i.i.d. Rayleigh fading channel model, we identify the effect of QoS limitations on the performance in the high-$\tsnr$ regime by determining the high-$\tsnr$ slope and power offset values.

\end{enumerate}

The organization of the rest of the paper is as follows. We describe the MIMO channel model in Section \ref{sec:model}. In Section \ref{sec:effcap}, we provide a description of the effective capacity formulation, and apply it to the MIMO setting. In Section \ref{sec:lowpower}, we study the effective capacity in the low-power regime and determine the first and second derivatives of the effective capacity at zero $\tsnr$. Subsequently, we apply the derivative expressions to investigate the energy efficiency. In Section \ref{sec:wideband}, we explore the effect of QoS limitations in the wideband regime, and identify the minimum bit energy requirements. In Section \ref{sec:highSNR}, we concentrate on the high-$\tsnr$ regime, and determine the impact of QoS constraints on the performance in the i.i.d. Rayleigh fading channel. Finally, we provide numerical results in Section \ref{sec:numerical} and conclude in Section \ref{sec:conclusion}.

\section{Channel Model} \label{sec:model}

We consider a MIMO channel model and assume that the transmitter and receiver are equipped with $n_T$ and $n_R$ antennas, respectively. Assuming flat-fading, we can express the channel input-output relation as
\begin{gather}
\y = \hh \x + \n. \label{eq:model}
\end{gather}
Above, $\x$ denotes the $n_T \times 1$--dimensional transmitted signal vector, and $\y$ represents the $n_R \times 1$--dimensional received signal vector. The channel input is assumed to be subject to the following average energy constraint:
\begin{gather} \label{eq:energyconstraint}
\E\{\|\x\|^2\} \le \frac{P}{B}
\end{gather}
where $B$ is the bandwidth of the system. When the bandwidth is $B$, we can assume that $B$ input vectors are transmitted every second, and \eqref{eq:energyconstraint} implies that the average power of the system is limited by $P$.
In (\ref{eq:model}),  $\n$ with dimension $n_R \times 1$ is a zero-mean Gaussian random vector with $E\{\n \n^\dd\} = N_0 \I$, where $\I$ is the identity matrix. The signal-to-noise ratio is defined as
\begin{gather} \label{eq:snr}
\tsnr = \frac{\E\{\|\x\|^2\}}{\E\{\|\n\|^2\}} = \frac{P}{n_R BN_0}.
\end{gather}
We also define the normalized input covariance matrix as
\begin{gather}
\K_x = \frac{\E\{\x \x^\dd\}}{P/B}.
\end{gather}
Note that the average energy constraint in \eqref{eq:energyconstraint} implies that the trace of the normalized covariance matrix is upper bounded by
\begin{gather}
\tr(\K_x) \le 1. \label{eq:traceconstraint}
\end{gather}
Finally, in \eqref{eq:model}, $\hh$ denotes the $n_R \times n_T$--dimensional random channel matrix whose components are the fading coefficients between the corresponding antennas at the transmitting and receiving ends. Unless specified otherwise, the components of $\hh$ are assumed to have arbitrary distributions with finite variances. Additionally, we consider the block-fading scenario and assume that the realization of the matrix $\hh$ remains fixed over a block of duration $T$ seconds and changes independently from one block to another.

\section{Effective Capacity of a MIMO Link} \label{sec:effcap}

In \cite{dapeng}, Wu and
Negi  defined the effective capacity as the maximum
constant arrival rate that a given service process can support in
order to guarantee a statistical QoS requirement specified by the
QoS exponent $\theta$ \footnote{For time-varying arrival rates,
effective capacity specifies the effective bandwidth of the arrival
process that can be supported by the channel.}. If we define $Q$ as
the stationary queue length, then $\theta$ is the decay rate of the
tail of the distribution of the queue length $Q$:
\begin{equation}
\lim_{q \to \infty} \frac{\log P(Q \ge q)}{q} = -\theta.
\end{equation}
Therefore, for large $q_{\max}$, we have the following approximation
for the buffer violation probability: $P(Q \ge q_{\max}) \approx
e^{-\theta q_{\max}}$. Hence, while larger $\theta$ corresponds to
more strict QoS constraints, smaller $\theta$ implies looser QoS
guarantees. Similarly, if $D$ denotes the steady-state delay
experienced in the buffer, then $P(D \ge d_{\max}) \approx
e^{-\theta \delta d_{\max}}$ for large $d_{\max}$, where $\delta$ is
determined by the arrival and service processes
\cite{Tang-crosslayer2}. Therefore, effective capacity formulation provides the maximum constant arrival rates that can be supported by the time-varying wireless channel under the queue length constraint $P(Q \ge q_{\max}) \le
e^{-\theta q_{max}}$ for large $q_{max}$ or the delay constraint $P(D \ge d_{\max}) \le
e^{-\theta \delta d_{\max}}$ for large $d_{\max}$. Since the average
arrival rate is equal to the average departure rate when the queue
is in steady-state \cite{ChangZajic}, effective capacity can also be
seen as the maximum throughput in the presence of such constraints.

The effective capacity is given by (\cite{dapeng}, \cite{chang}, \cite{cschang})
\begin{gather}
-\frac{\Lambda(-\theta)}{\theta}=-\lim_{t\rightarrow\infty}\frac{1}{\theta
t}\log_e{\mathbb{E}\{e^{-\theta S[t]}\}}
\end{gather}
where $S[t] = \sum_{i=1}^{t}R[i]$ is the time-accumulated service
process and $\{R[i], i=1,2,\ldots\}$ denotes the discrete-time
stationary and ergodic stochastic service process. Under the block-fading assumption, the effective capacity formulation simplifies to
\begin{gather} \label{eq:effcaporiginal}
-\frac{\Lambda(-\theta)}{\theta} = -\frac{1}{\theta T}\log_e\mathbb{E}\{e^{-\theta
T R[i]}\}.
\end{gather}
Under a short-term power constraint, the stochastic service process in a MIMO channel with a given normalized input covariance matrix $\K_x$ is
\begin{gather} \label{eq:servicerate}
B \log_2 \det \left(\I + \frac{P}{BN_0} \hh \K_x \hh^\dd\right) = B \log_2 \det \left(\I + n_R \tsnr \hh \K_x \hh^\dd\right) \,\, \text{bits/s}
\end{gather}
where $B$ denotes the bandwidth of the system and $\tsnr$ is as defined in (\ref{eq:snr}). We first consider the case in which $\hh$ is perfectly-known at the receiver and transmitter. In this scenario, the transmitter can adapt the input covariance matrix with respect to each realization of $\hh$ in order to maximize the service rate.
Therefore, using the formulation in (\ref{eq:effcaporiginal}), we can express the effective capacity normalized by the bandwidth and the receive dimensions as
\begin{gather}
\C_E(\tsnr, \theta) = -\frac{1}{\theta TB n_R}\log_e\mathbb{E}\left\{\exp\left(-\theta
T B\max_{\substack{\K_x \succeq \0 \\ \tr(\K_x) \le 1}} \log_2 \det \left(\I + n_R \tsnr \hh \K_x \hh^\dd\right)\right)\right\} \,\, \text{bits/s/Hz/dimension} \label{eq:effcapknown}
\end{gather}
As $\theta$ vanishes, the QoS constraints become loose and it can be easily verified that the effective capacity approaches the ergodic channel capacity, i.e.,
\begin{gather}
\lim_{\theta \to 0} \C_E(\tsnr,\theta) = \frac{1}{n_R}\E\left\{\max_{\substack{\K_x \succeq \0 \\ \tr(\K_x) \le 1}} \log_2 \det \left(\I + n_R \tsnr \hh \K_x \hh^\dd\right)\right\}.
\end{gather}
For $\theta > 0$, the effective capacity is in general smaller than the ergodic capacity. We can easily see this by interchanging the logarithm and the expectation in (\ref{eq:effcapknown}) and applying the Jensen's inequality:
\begin{align}
\C_E(\tsnr, \theta) &= -\frac{1}{\theta TB n_R}\log_e\mathbb{E}\left\{\exp\left(-\theta
T B\max_{\substack{\K_x \succeq \0 \\ \tr(\K_x) \le 1}} \log_2 \det \left(\I + n_R \tsnr \hh \K_x \hh^\dd\right)\right)\right\}
\\
&\le -\frac{1}{\theta TB n_R}\mathbb{E}\left\{\log_e\exp\left(-\theta
T B\max_{\substack{\K_x \succeq \0 \\ \tr(\K_x) \le 1}} \log_2 \det \left(\I + n_R \tsnr \hh \K_x \hh^\dd\right)\right)\right\}
\\
&= \frac{1}{n_R}\E\left\{\max_{\substack{\K_x \succeq \0 \\ \tr(\K_x) \le 1}} \log_2 \det \left(\I + n_R \tsnr \hh \K_x \hh^\dd\right)\right\}.
\end{align}

Above, we have assumed that $\hh$ is perfectly known at the transmitter. If, on the other hand, only statistical information regarding $\hh$ is available at the transmitter, then the input covariance matrix can be chosen to maximize the effective capacity. In such a case, the normalized effective capacity can be expressed as
\begin{gather}
\C_E(\tsnr, \theta) = \max_{\substack{\K_x \succeq \0 \\ \tr(\K_x) \le 1}}  -\frac{1}{\theta TB n_R}\log_e\mathbb{E}\left\{\exp\left(-\theta
T B \log_2 \det \left(\I + n_R \tsnr \hh \K_x \hh^\dd\right)\right)\right\} \,\, \text{bits/s/Hz/dimension}. \label{eq:effcapunknown}
\end{gather}
For a given (and not necessarily optimal) input covariance matrix $\K_x$, we call the throughput as effective rate and express it as
\begin{gather}
\R_E(\tsnr, \theta) = -\frac{1}{\theta TB n_R}\log_e\mathbb{E}\left\{\exp\left(-\theta
T B \log_2 \det \left(\I + n_R \tsnr \hh \K_x \hh^\dd\right)\right)\right\} \,\, \text{bits/s/Hz/dimension}. \label{eq:effrate}
\end{gather}
In practice, uniform power allocation across the antennas might be preferred. In this case, $\K_x = \frac{1}{n_T} \I$, and the effective rate can be written as
\begin{gather} \label{eq:effrateuniform}
\hspace{-.15cm}\R_{E, \text{id}}(\tsnr, \theta) =  -\frac{1}{\theta TB n_R}\log_e\mathbb{E}\left\{\exp\left(-\theta
T B \log_2 \det \left(\I + \frac{n_R}{n_T} \tsnr \hh  \hh^\dd\right)\right)\right\} \,\, \text{bits/s/Hz/dimension}
\end{gather}
where the subscript ``id" is introduced to denote that this expression is the throughput when the covariance matrix is proportional to an identity matrix.

Note that the effective capacity and effective rate expressions in (\ref{eq:effcapknown}), \eqref{eq:effcapunknown}, \eqref{eq:effrate}, and (\ref{eq:effrateuniform}) are proportional to the logarithm of the moment generating function of the instantaneous transmission rates.

Since the subsequent analysis assumes that the QoS exponent is fixed as power diminishes or increases or bandwidth increases, we generally suppress the argument $\theta$ and write the effective capacity and rate as $\C_E(\tsnr)$ and $\R_E(\tsnr)$, respectively.

Finally, before we go through a more detailed analysis of the effective capacity in the following sections, we would like to discuss several implicit assumptions made in the formulations provided in this section. The service rate expression in (\ref{eq:servicerate}) implies that the maximum transmission rates are equal to the instantaneous channel capacity in each block of duration $T$. Hence, we implicitly assume that the number of symbols in each block, $TB$, is large enough for this assumption to have operational meaning in practice. In (\ref{eq:effcapunknown}), it is assumed that the service rate is still given by (\ref{eq:servicerate}) and hence the transmitter employs variable-rate transmission scheme, even though the transmitter does not know the instantaneous realizations of $\hh$. Note this can be accomplished by using recently developed rateless codes such as LT \cite{Luby} or Raptor \cite{Shok} codes, which enable the transmitter to adapt its rate to the channel realization without requiring CSI at the transmitter side \cite{Castura1}, \cite{Castura2}. It is also important to note that the analysis conducted in this paper apply in the large-queue-length regime. If the buffer size is finite and small, then the arrival rates that can be supported by the system will be smaller than those considered in the paper, and in this case, one has to consider packet loss probabilities as well. Therefore, if the above-mentioned conditions and assumptions are not satisfied in the system, then the performance degradation will be more severe. For such cases, the results of this paper can be seen as fundamental limits (or upper bounds) which can serve as benchmarks for system performance.


\section{Effective Capacity in the Low-Power Regime} \label{sec:lowpower}

\subsection{First and Second Derivatives of the Effective Capacity}

In this section, we study the effective capacity in the low-$\tsnr$ regime and investigate the impact of the QoS exponent $\theta$. In particular, we consider the following second-order expansion of the effective capacity under different assumptions on the degree of channel state information:
\begin{gather}
\C_E(\tsnr) = \dot{\C}_E(0) \tsnr + \ddot{\C}_E(0) \frac{\tsnr^2}{2} + o(\tsnr^2)
\end{gather}
where $\dot{\C}_E(0)$ and $\ddot{\C}_E(0)$ denote the first and second derivatives of the effective capacity with respect to $\tsnr$ at $\tsnr = 0$. We first have the following result when the channel is perfectly known at the transmitter and receiver.

\begin{theo} \label{theo:derivsknown}
Assume that the realizations of the channel matrix $\hh$ are perfectly known at the receiver and transmitter. Assume further that the transmitter is subject to a short-term power constraint and hence is not allowed to perform power adaptation over time. Then, the first and second derivatives of the effective capacity in \eqref{eq:effcapknown} with respect to $\tsnr$ at $\tsnr = 0$ are
\begin{gather}
\dot{\C}_E(0) = \frac{1}{\log_e 2}\E\{\lm(\hh^\dd \hh)\} \label{eq:ecderiv}
\intertext{and}
\ddot{\C}_E(0) = \frac{\theta TB n_R}{\log_e^2 2}  \left[ \E^2\{\lm(\hh^\dd \hh)\} - \E\{\lm^2(\hh^\dd \hh)\} \right] - \frac{n_R}{l \log_e 2} \E\{\lm^2(\hh^\dd \hh)\} \label{eq:ecderiv2}
\end{gather}
where $\lm(\hh^\dd \hh)$ denotes the maximum eigenvalue of $\hh^\dd \hh$, and $l$ is the multiplicity of $\lm(\hh^\dd \hh)$.
\end{theo}

\emph{Proof}: For a given input covariance matrix $\K_x$, the effective rate is expressed as
\begin{align}
\R_{E}(\tsnr) &=  -\frac{1}{\theta TB n_R}\log_e\mathbb{E}\left\{\exp\left(-\theta
T B \log_2 \det \left(\I + n_R\tsnr \hh  \K_x \hh^\dd\right)\right)\right\} \label{eq:er1}
\\
&=-\frac{1}{\theta TB n_R}\log_e\mathbb{E}\left\{\exp\left(-\theta
T B \log_2 \det \left(\I + n_R\tsnr \bPhi\right)\right)\right\} \label{eq:er2}
\\
&=-\frac{1}{\theta TB n_R}\log_e\mathbb{E}\left\{\exp\left(-\theta
T B \sum_i \log_2 \left(1 + n_R\tsnr \lambda_i(\bPhi) \right)\right)\right\} \label{eq:er3}
\\
&=-\frac{1}{\theta TB n_R}\log_e\mathbb{E}\left\{\exp\left(-\frac{\theta
T B}{\log_e 2} \sum_i \log_e \left(1 + n_R\tsnr \lambda_i(\bPhi) \right)\right)\right\} \label{eq:er4}
\\
&=-\frac{1}{\theta TB n_R}\log_e\mathbb{E}\left\{f(\tsnr, \theta)\right\} \label{eq:er5}.
\end{align}
In \eqref{eq:er2} above, we have defined $\bPhi = \hh \K_x \hh^\dd$. \eqref{eq:er3} is obtained by noting that the determinant of a matrix is equal to the product of its eigenvalues, i.e., $\det \left(\I + n_R\tsnr \bPhi\right) = \prod_i (1 + n_R\tsnr \lambda_i(\bPhi))$, and also using the fact that the logarithm of a product is equal to the sum of the logarithms of the terms in the product. In \eqref{eq:er4}, the base of the logarithm is changed from 2 to $e$. In \eqref{eq:er5}, we have defined the function $f(\tsnr, \theta) = \exp\left(-\frac{\theta
T B}{\log_e 2} \sum_i \log_e \left(1 + n_R\tsnr \lambda_i(\bPhi) \right)\right)$.

Now, taking the derivative of $\R_E$ with respect to $\tsnr$ yields
\begin{align}
\dot{\R}_E(\tsnr) = -\frac{1}{\theta TB n_R}\frac{1}{\mathbb{E}\left\{f(\tsnr, \theta)\right\}} \E\left\{ -\frac{\theta TB}{\log_e 2} \sum_i \frac{n_r \lambda_i(\bPhi)}{1 + n_r \tsnr \lambda_i(\bPhi)} f(\tsnr, \theta)\right\}. \label{eq:erderiv}
\end{align}
Noting that the function $f$ evaluated at $\tsnr = 0$ is one, i.e., $f(0,\theta) = 1$, we can easily see from \eqref{eq:erderiv} that the value of the first derivative of the effective rate at $\tsnr = 0$ is
\begin{gather}
\dot{\R}_E(0) = \frac{1}{\log_e 2} \E \left\{ \sum_i \lambda_i(\bPhi) \right\} = \frac{1}{\log_e 2} \E \left\{ \tr(\bPhi) \right\} = \frac{1}{\log_e 2} \E \left\{ \tr(\hh \K_x \hh^\dd) \right\} \label{eq:erderivsnr0}
\end{gather}
where we have used the fact that the sum of the eigenvalues of a matrix is equal to its trace. Note that the normalized input covariance matrix $\K_x$ is by definition a positive semidefinite Hermitian matrix. As a Hermitian matrix, $\K_x$ can be written as \cite[Theorem 4.1.5]{matrixbook}
\begin{align} \label{eq:spectral}
\K_x = \U \bLambda \U^\dd = \sum_{i = 1}^{n_T} d_i \bu_i \bu_i^\dd
\end{align}
where $\U$ is a unitary matrix, $\{\bu_i\}$ are the column vectors of $\U$ and form an orthonormal set, $\bLambda$ is a real diagonal matrix, $\{d_i\}$ are the diagonal components of $\bLambda$. Since $\K_x$ is positive semidefinite, we have $d_i \ge 0$. Moreover, since all available energy should be used for transmission (i.e., the average energy and hence trace constraints should be satisfied with equality), we have $\tr(\K_x) = \sum_{i = 1}^{n_T}{d_i} = 1$. Combining (\ref{eq:erderivsnr0}) and (\ref{eq:spectral}), we can now write
\begin{align}
\dot{\R}_E(0) = \frac{1}{\log_e 2} \E \left\{ \tr(\hh \K_x \hh^\dd) \right\} &= \frac{1}{\log_e 2} \sum_{i = 1}^{n_T} d_i \E \left\{\tr( \hh \bu_i \bu_i^\dd \hh^\dd)\right\}
\\
&= \frac{1}{\log_e 2} \sum_{i = 1}^{n_T} d_i \E \left\{\bu_i^\dd \hh^\dd \hh \bu_i \right\} \label{eq:erderivsnr0-early}
\\
&\le \frac{1}{\log_e 2} \E\{\lm(\hh^\dd \hh)\}. \label{eq:erderivsnr0upper}
\end{align}
where $\lm(\hh^\dd \hh)$ denotes the maximum eigenvalue of the matrix $\hh^\dd \hh$.
The upper bound in (\ref{eq:erderivsnr0upper}) follows from the facts that $d_i \in [0,1]$ and $\sum_{i} d_i = 1$, and from
\cite[Theorem 4.2.2]{matrixbook} which states that since $\hh^\dd \hh$ is a Hermitian matrix and $\{\bu_i\}$ are unit vectors, we have
\begin{gather} \label{eq:upperbound-lambda}
\bu_i^\dd  \hh^\dd \hh \bu_i \le \lambda_{\max}(\hh^\dd \hh) \quad \forall i.
\end{gather}
The upper bound in (\ref{eq:erderivsnr0upper}) can be achieved by beamforming in the direction in which $\lm(\hh^\dd \hh)$ is achieved, i.e., by choosing the normalized input covariance matrix as
\begin{gather}
\K_x = \bu \bu^\dd
\end{gather}
where $\bu$ is the unit-norm eigenvector that corresponds to the maximum eigenvalue $\lm(\hh^\dd \hh)$. This lets us conclude that
\begin{gather}
\dot{\C}_E(0) = \frac{1}{\log_e 2} \E\{\lm(\hh^\dd \hh)\}
\end{gather}
proving \eqref{eq:ecderiv}.

Before proceeding to the proof of the second derivative result, we would like to note that transmission in the maximal-eigenvalue eigenspace  of $\hh^\dd \hh$ is indeed necessary to achieve the first derivative. Therefore, it is also necessary to attain the second derivative of the effective capacity at zero $\tsnr$. In a general scenario in which $\lm(\hh^\dd \hh)$ has a multiplicity of $l \ge 1$, an input covariance matrix in the following form is required:
\begin{gather} \label{eq:requiredcovariance}
\K_x = \sum_{i = 1}^l \alpha_i \bu_i \bu_i^\dd
\end{gather}
where $\alpha_i \in [0,1]$ and $\sum_{i =1}^l \alpha_i = 1$, and $\{\bu_i\}_{i=1}^l$ are the orthonormal eigenvectors that span the maximal-eigenvalue eigenspace of $\hh^\dd \hh$.

Now, we turn to the analysis of the second derivative. Differentiating $\dot{\R}_E$ in (\ref{eq:erderiv}) once more with respect to $\tsnr$, we obtain
\begin{align}\
\begin{split}
\ddot{\R}_E(\tsnr) = &-\frac{1}{\log_e 2}\frac{\E\left\{ -\frac{\theta TB}{\log_e 2} \sum_i \frac{n_r \lambda_i(\bPhi)}{1 + n_r \tsnr \lambda_i(\bPhi)} f(\tsnr, \theta)\right\}}{\E^2\left\{f(\tsnr, \theta)\right\}} \E\left\{ \sum_i \frac{\lambda_i(\bPhi)}{1 + n_r \tsnr \lambda_i(\bPhi)} f(\tsnr, \theta)\right\}
\\
&+ \frac{1}{\log_e2}\frac{1}{\E\left\{f(\tsnr, \theta)\right\}} \E\left\{ \sum_i \frac{-n_r \lambda_i^2(\bPhi)}{(1 + n_r \tsnr \lambda_i(\bPhi))^2} f(\tsnr, \theta)\right\}
\\
&-\frac{\theta T Bn_R}{\log_e^22}\frac{1}{\E\left\{f(\tsnr, \theta)\right\}} \E\left\{ \left(\sum_i \frac{ \lambda_i(\bPhi)}{1 + n_r \tsnr \lambda_i(\bPhi)}\right)^2 f(\tsnr, \theta)\right\}.
\end{split}\label{eq:erderiv2}
\end{align}
Again noting that $f(0, \theta) = 1$, we have
\begin{align}
\ddot{\R}_E(0) &= \frac{\theta T B n_R}{\log_e^22} \left(\E^2\left\{\sum_i \lambda_i(\bPhi)\right\} - \E\left\{\left( \sum_i \lambda_i(\bPhi)\right)^2  \right\}\right) - \frac{n_R}{\log_e2} \E\left\{\sum_i \lambda_i^2(\bPhi) \right\}
\\
&= \frac{\theta T B n_R}{\log_e^22} \left(\E^2\left\{\tr(\bPhi)\right\} - \E\left\{\tr^2( \bPhi)  \right\}\right) - \frac{n_R}{\log_e2} \E\left\{\tr(\bPhi^\dd\bPhi) \right\}. \label{eq:erderiv2snr0}
\end{align}
In obtaining (\ref{eq:erderiv2snr0}), we have used the facts that $\sum_i \lambda_i(\bPhi) = \tr(\bPhi)$ and $\sum_i \lambda_i^2(\bPhi) = \tr(\bPhi^\dd \bPhi)$.

As described above, an input covariance matrix that is in the form given in (\ref{eq:requiredcovariance}) is required to achieve the second derivative of the effective capacity at $\tsnr = 0$. For such a covariance matrix, it can be easily verified that
\begin{gather}
\E\left\{\tr(\bPhi)\right\} = \E\left\{\tr(\hh \K_x \hh^\dd)\right\} = \E\left\{\lm(\hh^\dd \hh)\right\} \label{eq:tracebPhi}
\end{gather}
and
\begin{align}
\E\left\{\tr(\bPhi^\dd\bPhi) \right\} = \E\left\{\tr(\hh \K_x \hh^\dd\hh \K_x \hh^\dd)\right\} &= \E\left\{\sum_{i,j}^l \alpha_i \alpha_j |\bu_j^\dd \hh^\dd \hh \bu_i|^2\right\} \label{eq:traceeqn1}
\\
&=\E\left\{\lm^2(\hh^\dd \hh)\sum_{i,j}^l \alpha_i \alpha_j |\bu_j^\dd \bu_i|^2\right\} \label{eq:traceeqn2}
\\
&=\E\left\{\lm^2(\hh^\dd \hh)\sum_{i=1}^l \alpha_i^2\right\} \label{eq:traceeqn3}
\\
&\ge\frac{1}{l}\E\left\{\lm^2(\hh^\dd \hh)\right\} \label{eq:traceeqn4}
\end{align}
where (\ref{eq:traceeqn2}) follows from the fact that $\{\bu_i\}$ are the eigenvectors that correspond to $\lm(\hh^\dd \hh)$ and hence $\hh^\dd \hh \bu_i = \lm(\hh^\dd \hh) \bu_i$, (\ref{eq:traceeqn3}) follows from the orthonormality of $\{\bu_i\}$ which implies that
\begin{gather}
\bu_j^\dd \bu_i = \left\{
\begin{array}{ll}
1 & \text{if } i =j
\\
0 & \text{if } i \neq j
\end{array}
\right..
\end{gather}
Finally, (\ref{eq:traceeqn4}) follows from the properties that $\alpha_i \in [0,1]$ and $\sum_{i=1}^l \alpha_i = 1$, and the fact that $\sum_{i=1}^l \alpha_i^2$ under these properties is minimized by choosing $\alpha_i = \frac{1}{l}$, which leads to the lower bound $\sum_{i=1}^l \alpha_i^2 \ge \frac{1}{l}$.

We note from (\ref{eq:tracebPhi}) that given the required the covariance structure in (\ref{eq:requiredcovariance}), the first term in the expression of $\ddot{\R}_E(0)$ in \eqref{eq:erderiv2snr0} is $\frac{\theta T B n_R}{\log_e^22} \left(\E^2\left\{\tr(\bPhi)\right\} - \E\left\{\tr^2( \bPhi)  \right\}\right) =  \frac{\theta T B n_R}{\log_e^22} \left(\E^2\left\{\lm(\hh^\dd \hh)\right\} - \E\left\{\lm^2(\hh^\dd \hh)\right\}\right)$ for all possible $\{\alpha_i\}$. On the other hand, the second term in (\ref{eq:erderiv2snr0}) is minimized by having $\alpha_i = \frac{1}{l}$ for all $i$, i.e., by equally allocating the power in the orthogonal directions in the maximal-eigenvalue eigenspace. Therefore, the input covariance matrix $\K_x = \frac{1}{l }\sum_{i = 1}^l \bu_i \bu_i^\dd$ maximizes $\ddot{\R}_E(0)$, and we have
\begin{gather}
\ddot{\C}_E(0) = \frac{\theta T B n_R}{\log_e^22} \left(\E^2\left\{\lm(\hh^\dd \hh)\right\} - \E\left\{\lm^2(\hh^\dd \hh)\right\}\right) - \frac{n_R}{l \log_e2} \E\left\{\lm^2(\hh^\dd \hh)\right\}
\end{gather}
proving \eqref{eq:ecderiv2}. \hfill $\blacksquare$

Next, we consider the case in which the transmitter has only statistical knowledge of the channel.

\begin{theo} \label{theo:derivsunknown}
Assume that while the receiver perfectly knows the channel matrix $\hh$, the transmitter only has the knowledge of $\E\{\hh^\dd \hh\}$. Then, the first and second derivatives of the effective capacity in (\ref{eq:effcapunknown}) are
\begin{gather}
\dot{\C}_E(0) = \frac{1}{\log_e 2}\lm(\E\{\hh^\dd \hh\}) \label{eq:ecderivunknown}
\intertext{and}
\!\!\!\!\!\!\!\ddot{\C}_E(0) = \frac{\theta TB n_R}{\log_e^2 2} \lm^2(\E\{\hh^\dd \hh\}) - \!\!\!\!\min_{\substack{\{\alpha_i\} \\ \alpha_i \in [0,1] \, \forall i \\ \sum_{i=1}^l \alpha_i = 1}} \sum_{i,j}^{l} \alpha_i \alpha_j \left(  \frac{\theta TB n_R}{\log_e^2 2} \E\{(\bu_i^\dd \hh^\dd \hh \bu_i)(\bu_j^\dd \hh^\dd \hh \bu_j)\} + \frac{n_R}{\log_e 2} \E\{|\bu_j^\dd \hh^\dd \hh \bu_i|^2\} \right) \label{eq:ecderiv2unknown}
\end{gather}
where $\lm(\E\{\hh^\dd \hh\})$ denotes the maximum eigenvalue of $\E\{\hh^\dd \hh\}$, and $l$ is the multiplicity of $\lm(E\{\hh^\dd \hh\})$.
\end{theo}

\emph{Proof}: Note from \eqref{eq:erderivsnr0-early} that for a given covariance matrix $\K_x = \U \bLambda \U^\dd = \sum_{i = 1}^{n_T} d_i \bu_i \bu_i^\dd$, the first derivative of the effective rate is
\begin{align}
\dot{\R}_E(0) &= \frac{1}{\log_e 2} \sum_{i = 1}^{n_T} d_i \E \left\{\bu_i^\dd \hh^\dd \hh \bu_i \right\} \label{eq:erderivsnr0unknown1}
\\
&= \frac{1}{\log_e 2} \sum_{i = 1}^{n_T} d_i \bu_i^\dd  \E \left\{\hh^\dd \hh \right\} \bu_i \label{eq:erderivsnr0unknown2}
\\
&\le \frac{1}{\log_e 2} \lm(\E\{\hh^\dd \hh\}) \label{eq:erderivsnr0unknown3}
\end{align}
where (\ref{eq:erderivsnr0unknown2}) follows by noting that the transmitter has only statistical knowledge of $\hh$, and the input covariance matrix and hence $\{\bu_i\}$ cannot depend on the realizations of $\hh$. Therefore, $\{\bu_i\}$ are deterministic and can be taken out of the expectation. Now, the upper bound in (\ref{eq:erderivsnr0unknown3}), similarly as discussed in the proof of Theorem \ref{theo:derivsknown}, is achieved by transmitting in the maximal-eigenvalue eigenspace of $\E\{\hh^\dd \hh\}$. Therefore, a covariance matrix in the following form is required to achieve the first derivative of the effective capacity:
\begin{gather}
\K_x = \sum_{i = 1}^{l} \alpha_i \bu_i \bu_i^\dd \label{eq:requiredcovarianceunknown}
\end{gather}
where $\{\bu_i\}$ are the orthonormal eigenvectors spanning the maximal-eigenvalue eigenspace of $\E\{\hh^\dd \hh\}$, $l$ is the multiplicity of $\lm(\E\{\hh^\dd \hh\})$, and $\{\alpha_i\}$ are constants taking values in $[0,1]$ and has unit sum, i.e., $\sum_{i=1}^l \alpha_i = 1$. Consequently, this covariance structure is also necessary to attain the second derivative of the effective capacity. Employing the second derivative expression in (\ref{eq:erderiv2snr0}) with the covariance matrix in (\ref{eq:requiredcovarianceunknown}), and maximizing $\dot{\R}_E(0)$ with respect to all possible choices of $\{\alpha_i\}$, we easily obtain (\ref{eq:ecderiv2unknown}).
\hfill $\blacksquare$

Using the results seen in the proofs of Theorems \ref{theo:derivsknown} and \ref{theo:derivsunknown}, we can also immediately obtain the following result when the power is uniformly distributed across the transmit antennas and hence we have $\K_x = \frac{1}{n_T} \I$.

\begin{cor} \label{cor:derivsuniform}
Assume that the input covariance matrix is $\K_x = \frac{1}{n_T} \I$. Then, the first and second derivatives of the effective rate $\R_{E, \text{id}}$ given in (\ref{eq:effrateuniform}) are
\begin{gather}
\dot{\R}_{E,\text{id}}(0) = \frac{1}{n_T\log_e 2}\E\{\tr(\hh^\dd \hh)\} \label{eq:erderivuniform}
\intertext{and}
\ddot{\R}_{E,\text{id}}(0) = \frac{\theta TB n_R}{n_T^2\log_e^2 2}  \left[ \E^2\{\tr(\hh^\dd \hh)\} - \E\{\tr^2(\hh^\dd \hh)\} \right] - \frac{n_R}{n_T^2 \log_e 2} \E\{\tr((\hh^\dd \hh)^2)\}. \label{eq:erderiv2uniform}
\end{gather}
\end{cor}

\begin{rem}
Note that the common theme in the results of Theorems \ref{theo:derivsknown} and \ref{theo:derivsunknown}, and Corollary \ref{cor:derivsuniform} is that the first derivative does not depend on $\theta$ and hence does not get affected by the presence of QoS constraints. Indeed, the first derivative expressions are equal to the ones obtained when Shannon capacity, rather than effective capacity, is considered \cite{verdu}. On the other hand, the second derivative is a function of $\theta$ and in general decreases as $\theta$ increases or equivalently as the queueing constraints become more stringent\footnote{Note that $\E^2\{\lm(\hh^\dd \hh)\} \le \E\{\lm^2(\hh^\dd \hh)\}$ and $\E^2\{\tr(\hh^\dd \hh)\} \le \E\{\tr^2(\hh^\dd \hh)\}$.}.
\end{rem}

\subsection{Energy Efficiency in the Low-Power Regime} \label{subsec:energy}

The expressions of the first and second derivatives enable us to analyze the energy efficiency in the low-power regime. The minimum bit energy under QoS constraints is given by \cite{verdu}
\begin{equation}\label{ebmin}
\frac{E_b}{N_0}_{\text{min}}=\lim_{\tsnr\rightarrow0}\frac{\tsnr}{\C_{E}(\tsnr)}=\frac{1}{\dot{\C}_{E}(0)}.
\end{equation}
At $\frac{E_b}{N_0}_{\text{min}}$, the slope $\mathcal {S}_0$ of the
spectral efficiency versus $E_b/N_0$ (in dB) curve is defined as
\cite{verdu}
\begin{equation}\label{slope}
\mathcal{S}_0=\lim_{\frac{E_b}{N_0}\downarrow\frac{E_b}{N_0}_\text{min}}
\frac{\C_E(\frac{E_b}{N_0})}{10\log_{10}\frac{E_b}{N_0}-10\log_{10}\frac{E_b}{N_0}_\text{min}}10\log_{10}2.
\end{equation}
Considering the expression for normalized effective capacity, the
wideband slope can be found from \cite{verdu}\footnote{We note that the expressions in (\ref{ebmin}) and (\ref{ecslope}) differ from those in \cite{verdu} by a constant factor due to our assumption that the units of $\C_E$ is bits/s/Hz/dimension rather than nats/s/Hz/dimension.}
\begin{equation}\label{ecslope}
\mathcal{S}_0=\frac{2(\dot{\C}_E(0))^2}{-\ddot{\C}_E(0)}\log_e{2} \quad \text{bits/s/Hz/(3 dB)/receive antenna}.
\end{equation}
\begin{cor}
Applying the results of Theorem \ref{theo:derivsknown} to the above formulation, we obtain
\begin{gather}\label{eq:minenergyQoS}
\frac{E_b}{N_0}_{\min} = \frac{\log_e 2}{\E\{\lm(\hh^\dd \hh)\}}
\end{gather}
\begin{align}
\mathcal{S}_0&=\frac{2\E^2\{\lm(\hh^\dd \hh)\}}{\frac{n_R}{l}\E\{\lm^2(\hh^\dd \hh)\} + \frac{\theta TBn_R}{\log_e2} \left(\E\{\lm^2(\hh^\dd \hh)\} - \E^2\{\lm(\hh^\dd \hh)\}\right)}
\\
&=\frac{2}{\frac{n_R}{l}\kappa(\sigma_{\max}(\hh)) + \frac{\theta TBn_R}{\log_e2} \left(\kappa(\sigma_{\max}(\hh))-1\right)} \label{eq:widebandslopeQoS}
\end{align}
where $\kappa(\sigma_{\max}(H))$ is the kurtosis of maximum singular value of the matrix $\hh$ and is defined as
\begin{gather}
\kappa(\sigma_{\max}(\hh)) = \frac{\E\{\sigma_{\max}^4(\hh)\}}{\E^2\{\sigma^2_{\max}(\hh)\}} = \frac{\E\{\lm^2(\hh^\dd \hh)\}}{\E^2\{\lm(\hh^\dd \hh)\}}.
\end{gather}
\end{cor}
\begin{rem}
In \cite{verdu}, Shannon capacity is considered and it is shown that $\frac{E_b}{N_0}_{\min} = \frac{\log_e 2}{\E\{\lm(\hh^\dd \hh)\}}$ and $\mathcal{S}_0 = \frac{2}{\frac{n_R}{l}\kappa(\sigma_{\max}(\hh))}$. From (\ref{eq:minenergyQoS}) and (\ref{eq:widebandslopeQoS}) we note that we have the same minimum bit energy in the presence of QoS limitations while the wideband slope diminishes with increasing $\theta$.
\end{rem}

When we have equal power allocation, i.e., $\K_x = \frac{1}{n_T} \I$, it can be immediately seen from the result of Corollary \ref{cor:derivsuniform} that
\begin{gather}
\frac{E_b}{N_0}_{\min} = \frac{n_T\log_e 2}{\E\{\tr(\hh^\dd \hh)\}}
\\
\mathcal{S}_0=\frac{2\E^2\{\tr(\hh^\dd \hh)\}}{n_R\E\{\tr((\hh^\dd \hh)^2)\} + \frac{\theta TBn_R}{\log_e2} \left(\E\{\tr^2(\hh^\dd \hh)\} - \E^2\{\tr(\hh^\dd \hh)\}\right)}.
\end{gather}
Assume that $\hh$ has independent zero-mean unit-variance complex Gaussian random entries. Under this assumption, we have \cite{lozano}
\begin{align}
\E\{\tr(\hh^\dd \hh)\} = n_Rn_T, \quad \E\{\tr^2(\hh^\dd \hh)\}=n_Rn_T(n_Rn_T+1), \quad \E\{\tr((\hh^\dd \hh)^2)\} = n_Rn_T(n_R + n_T).
\end{align}
Using these facts, we have the following minimum bit energy and wideband slope expressions for the uniform power allocation case when the entries of $\hh$ are i.i.d. zero-mean unit-variance Gaussian random variables:
\begin{align} \label{eq:minenergy-iidRayleigh}
\frac{E_b}{N_0}_{\min} = \frac{\log_e 2}{n_R}
\quad \text{and} \quad
\mathcal{S}_0 = \frac{2}{\frac{n_R + n_T}{n_T} + \frac{\theta TB}{n_T \log_e2}} \quad \text{bits/s/Hz/(3 dB)/receive antenna}.
\end{align}
We note that while the minimum bit energy depends only on the number of receive antennas, the wideband slope is a function of both the receive and transmit antennas. Note that the wideband slope expression is per receive antenna. Without this normalization, we have
\begin{gather} \label{eq:widebandslopeunnormalized}
\mathcal{S}_0 = \frac{2}{\frac{n_R + n_T}{n_r n_T} + \frac{\theta TB}{n_R n_T \log_e2}} \quad \text{bits/s/Hz/(3 dB)}.
\end{gather}
From (\ref{eq:widebandslopeunnormalized}), we identify the interactions between the spatial dimensions and QoS constraints. Note that more strict QoS constraints and hence higher values of $\theta$ tend to diminish the wideband slope. On the other hand, we see in the second term in the denominator of (\ref{eq:widebandslopeunnormalized}) that the impact of the presence of QoS constraints is being diminished by the product of the number of transmit and receive antennas, $n_R n_T$. Hence, increasing the number of transmit and/or receive antennas can offset the performance loss due to queueing constraints.

\section{Minimum Bit Energy in the Wideband Regime} \label{sec:wideband}

In the previous section, we have assumed that the bandwidth of the system is fixed as the transmission power $P$ diminishes and system operates in the low-power regime. Here, we study the regime in which the bandwidth increases while $P$ is kept fixed. Note that as the bandwidth grows, the flat-fading assumption will no longer hold and the input-output relation given in (\ref{eq:model}) will not be an accurate description. On the other hand, if we decompose the wideband channel into parallel, noninteracting, narrowband subchannels each with bandwidth that is equal to the coherence bandwidth $B_c$, then we can assume that each subchannel experiences independent flat fading and has an input-output relation that can be expressed as
\begin{gather}\label{eq:modelwb}
\y_i = \hh_i \x_i + \n_i \quad i = 1,2,\ldots, m
\end{gather}
where $\x_i$ and $\y_i$ are the input and output vectors of the $i^{\text{th}}$ subchannel, and $\hh_i$ is the $i^{\text{th}}$ subchannel matrix. $\n_i$ represents the additive zero mean Gaussian noise vector with $E\{\n_i \n_i^\dd\} = N_0 \I$ in the  $i^{\text{th}}$ subchannel. We assume that the input in the $i^{\text{th}}$ subchannel is subject to $E\{\|\x_i\|^2\} \le \frac{P_i}{B_c}$ where $P_i$ is the power allocated to the $i^{\text{th}}$ subchannel. We assume that the number of subchannels is $m$ and hence we have $B = mB_c$ and $\sum_{i=1}^m P_i = P$ where $B$ and $P$ denote the total bandwidth and power, respectively, of the wideband system. Under these assumptions, the maximum instantaneous transmission rate in the $i^{\text{th}}$ subchannel with covariance matrix $\K_{x,i}$ is
\begin{gather}
B_c \log_2 \det \left(\I + n_R \tsnr_i \hh_i \K_{x,i} \hh_i^\dd\right) \,\, \text{bits/s}
\end{gather}
where $\tsnr_i = \frac{P_i}{n_R B_c N_0}$.
Due to the independence of fading in different subchannels, the total transmission rate over the wideband channel is
\begin{gather}\label{eq:totaltransmissionrate}
\sum_{i=1}^m B_c \log_2 \det \left(\I + n_R \tsnr_i \hh_i \K_{x,i} \hh_i^\dd\right) \,\, \text{bits/s}
\end{gather}
which is achieved by independent signaling over different subchannels, i.e., by choosing $\{\x_i\}_{i=1}^m$ as zero-mean independent Gaussian vectors with covariance matrices $\{\K_{x,i}\}_{i=1}^m$. Then, for the transmission rate in (\ref{eq:totaltransmissionrate}), the effective rate is given by
\begin{align}
\R_E(\tsnr) &= -\frac{1}{\theta TB n_R}\log_e\mathbb{E}\left\{\exp\left(-\theta
T B_c \sum_{i=1}^m \log_2 \det \left(\I + n_R \tsnr_i \hh_i \K_{x,i} \hh_i^\dd\right)\right)\right\} 
\\
&=-\frac{1}{\theta TB n_R}\log_e \prod_{i=1}^m \mathbb{E}\left\{ \exp\left(-\theta
T B_c \log_2 \det \left(\I + n_R \tsnr_i \hh_i \K_{x,i} \hh_i^\dd\right)\right)\right\} \label{eq:effratewb1}
\\
&=-\frac{1}{\theta TB n_R} \sum_{i=1}^m\log_e  \mathbb{E}\left\{ \exp\left(-\theta
T B_c \log_2 \det \left(\I + n_R \tsnr_i \hh_i \K_{x,i} \hh_i^\dd\right)\right)\right\} \label{eq:effratewb2}
\end{align}
where (\ref{eq:effratewb1}) follows from our assumption that $\{\hh_i\}$ are independent subchannel matrices and the fact that the expected value of a product of independent random variables is equal to the product of the expected values of the individual random variables. In general, effective capacity can be obtained by maximizing the effective rate expression in (\ref{eq:effratewb2}) over all power allocations $\{P_i\}$ and covariance matrices $\{\K_{x,i}\}$. If the channel is known at the transmitter, $\{P_i\}$ and $\{\K_{x,i}\}$ can depend on the realizations of the channel matrices $\{\hh_i\}$.

We simplify the above setting by assuming that $\hh_i \K_{x,i} \hh_i^\dd$ has the same distribution for all $i = 1,2,\ldots, m$. For instance, this assumption would hold when $\{\hh_i\}$ are identically distributed, and $\K_{x,i}$ is the same fixed matrix for all $i$ or is a random matrix with a common distribution for all $i$ (e.g., $\K_{x,i} = \bu\bu^\dd$, where $\bu$ is the random eigenvector that corresponds to $\lm(\hh_i^\dd \hh_i) $, has the same distribution for all $i$ when $\{\hh_i\}$ are identically distributed). Under this assumption, we can eliminate the dependence of $\hh_i \K_{x,i} \hh_i^\dd$ on the time index $i$, and show from the concavity of the expression (\ref{eq:effratewb2}) with respect to signal-to-noise ratio\footnote{Since $-\theta
T B_c \log_2 \det \left(\I + n_R \tsnr_i \hh_i \K_{x,i} \hh_i^\dd\right)$ is a convex function of $\tsnr$ for given $\hh_i \K_{x,i} \hh_i^\dd$, $e^{-\theta
T B_c \log_2 \det \left(\I + n_R \tsnr_i \hh_i \K_{x,i} \hh_i^\dd\right)}$ is a log-convex function. Moreover, since log-convexity is preserved under sums \cite[Section 3.5.2]{convexitybook}, $\mathbb{E}\left\{ \exp\left(-\theta
T B_c \log_2 \det \left(\I + n_R \tsnr_i \hh_i \K_{x,i} \hh_i^\dd\right)\right)\right\}$ is log-convex, implying that $\log_e \mathbb{E}\left\{ \exp\left(-\theta
T B_c \log_2 \det \left(\I + n_R \tsnr_i \hh_i \K_{x,i} \hh_i^\dd\right)\right)\right\}$ is a convex function of $\tsnr$. Since the sum of convex functions is convex \cite{convexitybook}, and the negative of a convex function is concave, we conclude that the expression in (\ref{eq:effratewb2}) is a concave function of $\tsnr$.} that the effective rate is maximized by having $\tsnr_i = \frac{P/m}{n_R N_0 B_c} = \frac{P}{n_R N_0 B} = \tsnr$ for all $i$, i.e., by distributing the total power equally over the subchannels. Now, the effective rate expression becomes
\begin{align}
\R_E(\tsnr)&=-\frac{1}{\theta TB n_R} m \log_e  \mathbb{E}\left\{ \exp\left(-\theta
T B_c \log_2 \det \left(\I + n_R \tsnr \hh \K_{x} \hh^\dd\right)\right)\right\}
\\
&=-\frac{1}{\theta TB_c n_R}\log_e  \mathbb{E}\left\{ \exp\left(-\theta
T B_c \log_2 \det \left(\I + n_R \tsnr \hh \K_{x} \hh^\dd\right)\right)\right\} \label{eq:effratewbsimplified}
\end{align}
where we have used the relation $B = mB_c$.

Now, we analyze the effective capacity and energy efficiency in the wideband limit in three scenarios:

\subsubsection{Rich Multipath Fading}
In a system with bandwidth $B$, the maximum number of resolvable paths is proportional to $B T_m = \frac{B}{B_c}$ where $T_m$ denotes the delay spread and $B_c = \frac{1}{T_m}$.
In rich multipath fading, the assumption is that the number of independent resolvable paths increases linearly with increasing bandwidth.  Therefore, in rich multipath fading, coherence bandwidth $B_c$ remains fixed as $B$ increases while $\tsnr = \frac{P}{BN_0}$ diminishes to zero. Then, from the similarity of the effective rate expressions in (\ref{eq:effrate}) and \eqref{eq:effratewbsimplified} and the fact that $B$ is fixed in (\ref{eq:effrate}) in the low-power regime analysis, we immediately conclude that the wideband and low-power results are identical in rich multipath fading under the assumptions that lead to the effective rate expression in (\ref{eq:effratewbsimplified}).

\subsubsection{Sparse Multipath Fading}

In sparse multipath fading, it is assumed that the number of independent resolvable paths increases at most \emph{sublinearly} with bandwidth \cite{porrat} \cite{raghavan}. Hence, in this case, $B_c$ increases with increasing bandwidth. In the special case in which the number of resolvable paths is bounded, $B_c$ increases linearly with $B$ while the number of subchannels $m$ remains fixed. For instance, such a scenario is considered in \cite{telatar}. For this case, we have the following result on the minimum bit energy required in the wideband regime.

\begin{theo} \label{theo:minbitenergywb}
Assume that the number of independent resolvable paths remain bounded and fixed in the wideband regime as $B$ increases. In this case, the minimum bit energy for a given covariance matrix $\K_x$ is given by
\begin{gather} \label{eq:minbitenergywbKx}
\frac{E_b}{N_0}_{\min} = \frac{\frac{\theta T P}{mN_0}}{-\log_e \E\left\{e^{-\frac{\theta T P}{mN_0} \frac{1}{\log_e 2} \, \tr(\hh \K_x \hh^\dd)}\right\}}.
\end{gather}
When the channel is perfectly known at the transmitter, information can be sent in the maximal-eigenvalue eigenspace of $\hh^\dd \hh$ and the required minimum bit energy becomes
\begin{gather}\label{eq:minbitenergywbknown}
\frac{E_b}{N_0}_{\min} = \frac{\frac{\theta T P}{mN_0}}{-\log_e \E\left\{e^{-\frac{\theta T P}{mN_0} \frac{1}{\log_e 2} \, \lm(\hh^\dd \hh)}\right\}}.
\end{gather}
If only statistical information of the channel is available at the transmitter, the minimum bit energy can be obtained by minimizing (\ref{eq:minbitenergywbKx}) over all permissible covariance matrices, i.e.,
\begin{gather} \label{eq:minbitenergywbunknown}
\frac{E_b}{N_0}_{\min} = \min_{\substack{\K_x \succeq \0 \\ \tr(\K_x) \le 1}}\frac{\frac{\theta T P}{mN_0}}{-\log_e \E\left\{e^{-\frac{\theta T P}{mN_0} \frac{1}{\log_e 2} \, \tr(\hh \K_x \hh^\dd)}\right\}}.
\end{gather}
\end{theo}
\emph{Proof:} For a given input covariance matrix $\K_x$, the bit energy required for reliable communications under QoS constraints is
\begin{align}
\frac{E_b}{N_0} = \frac{\tsnr}{\R_E(\tsnr)} &= \frac{\frac{P}{n_R B N_0}}{-\frac{1}{\theta TB_c n_R}\log_e  \mathbb{E}\left\{ \exp\left(-\theta
T B_c \log_2 \det \left(\I + n_R \tsnr \hh \K_{x} \hh^\dd\right)\right)\right\}}
\\
&= \frac{\frac{\theta TP}{m N_0}}{-\log_e  \mathbb{E}\left\{ \exp\left(-\theta
T B_c \log_2 \det \left(\I + \frac{P}{m B_c N_0} \hh \K_{x} \hh^\dd\right)\right)\right\}} \label{eq:ebnowb1}
\\
&= \frac{\frac{\theta TP}{m N_0}}{-\log_e  \mathbb{E}\left\{ \exp\left(-\theta
T B_c \sum_i \log_2 \left(1 + \frac{P}{m B_c N_0} \lambda_i(\hh \K_{x} \hh^\dd)\right)\right)\right\}} \label{eq:ebnowb2}
\end{align}
where $\lambda_i(\hh \K_{x} \hh^\dd)$ denotes the $i^{\text{th}}$ eigenvalue of the matrix $\hh \K_{x} \hh^\dd$. Above, (\ref{eq:ebnowb1}) is obtained by using the relation $B = m B_c$ and performing some straightforward algebraic operations, and (\ref{eq:ebnowb2}) follows from the fact that $\det(\mathbf{A}) = \prod_i \lambda_i(\mathbf{A})$. Note that under the assumption of fixed number of resolvable paths, $B_c$ increases linearly with $B$ while $m$ is fixed. Hence, only the denominator of (\ref{eq:ebnowb2}) varies with $B$. From the fact that the function $x \log_2 (1 + \frac{a}{x})$ is a monotonically increasing function of $x > 0$ for any constant $a > 0$, we can easily see that the minimum bit energy is achieved as $B \to \infty$. Since $B_c$ also grows without  bound as $B$ increases, we have
\begin{align}
\frac{E_b}{N_0}_{\min} &= \lim_{B_c \to \infty} \frac{\frac{\theta TP}{m N_0}}{-\log_e  \mathbb{E}\left\{ \exp\left(-\theta
T B_c \sum_i \log_2 \left(1 + \frac{P}{m B_c N_0} \lambda_i(\hh \K_{x} \hh^\dd)\right)\right)\right\}}
\\
&= \frac{\frac{\theta TP}{m N_0}}{-\log_e  \mathbb{E}\left\{ \exp\left(-\theta
T \frac{1}{\log_e 2} \sum_i  \frac{P}{m N_0} \lambda_i(\hh \K_{x} \hh^\dd)\right)\right\}} \label{eq:ebnominwb1}
\\
&= \frac{\frac{\theta TP}{m N_0}}{-\log_e  \mathbb{E}\left\{ \exp\left(-\frac{\theta
T P}{m N_0} \frac{1}{\log_e 2} \sum_i  \lambda_i(\hh \K_{x} \hh^\dd)\right)\right\}} \label{eq:ebnominwb2}
\\
&= \frac{\frac{\theta TP}{m N_0}}{-\log_e  \mathbb{E}\left\{ \exp\left(-\frac{\theta
T P}{m N_0} \frac{1}{\log_e 2} \, \tr(\hh \K_{x} \hh^\dd)\right)\right\}}.\label{eq:ebnominwb3}
\end{align}
(\ref{eq:ebnominwb1}) is obtained using the fact that as $B_c \to \infty$, we have $B_c \log_2 \left(1 + \frac{P}{m B_c N_0} \lambda_i(\hh \K_{x} \hh^\dd)\right) \to \frac{1}{\log_e 2}\frac{P}{mN_0} \lambda_i(\hh \K_{x} \hh^\dd)$. (\ref{eq:ebnominwb3}) follows from the property that $\sum_i \lambda_i(\mathbf{A}) = \tr(\mathbf{A})$. Note that (\ref{eq:ebnominwb3}) proves (\ref{eq:minbitenergywbKx}) which is the minimum bit energy for a given covariance matrix $\K_x$.

Recall that it is shown in the proof of Theorem \ref{theo:derivsknown} that
\begin{gather}
\tr(\hh\K_x \hh^\dd) \le \lm(\hh^\dd \hh)
\end{gather}
and this upper bound can be achieved by transmitting in the maximal-eigenvalue eigenspace of $\hh\hh^\dd$, e.g., by having $\K_x = \bu \bu^\dd$ where $\bu$ is the eigenvector that corresponds to $\lm(\hh^\dd \hh)$. If the transmitter perfectly knows the realizations of the channel matrix $\hh$, then this transmission strategy can be employed and the minimum bit energy becomes
\begin{gather}
\frac{E_b}{N_0}_{\min} = \frac{\frac{\theta TP}{m N_0}}{-\log_e  \mathbb{E}\left\{ \exp\left(-\frac{\theta
T P}{m N_0} \frac{1}{\log_e 2} \, \lm(\hh  \hh^\dd)\right)\right\}}.
\end{gather}
If the transmitter has only statistical knowledge of the channel matrix, the minimum bit energy can be determined by finding the input covariance matrix that minimizes (\ref{eq:ebnominwb3}). \hfill $\blacksquare$

\begin{rem} \label{rem:increasedenergy}
By applying the Jensen's inequality, we can easily see that
\begin{gather}
\log_e \E\left\{e^{-\frac{\theta T P}{mN_0} \frac{1}{\log_e 2} \, \tr(\hh \K_x \hh^\dd)}\right\} \ge \E\left\{ \log_e e^{-\frac{\theta T P}{mN_0} \frac{1}{\log_e 2} \, \tr(\hh \K_x \hh^\dd)}\right\} = \E\left\{-\frac{\theta T P}{mN_0} \frac{1}{\log_e 2} \, \tr(\hh \K_x \hh^\dd)\right\}
\end{gather}
which implies that
\begin{gather}
\frac{E_b}{N_0}_{\min} = \frac{\frac{\theta T P}{mN_0}}{-\log_e \E\left\{e^{-\frac{\theta T P}{mN_0} \frac{1}{\log_e 2} \, \tr(\hh \K_x \hh^\dd)}\right\}} \ge \frac{\log_e2}{\tr(\hh \K_x \hh^\dd)}.
\end{gather}
Similarly, we can show
\begin{gather}
\frac{E_b}{N_0}_{\min} = \frac{\frac{\theta T P}{mN_0}}{-\log_e \E\left\{e^{-\frac{\theta T P}{mN_0} \frac{1}{\log_e 2} \, \lm(\hh^\dd \hh)}\right\}} \ge \frac{\log_e2}{\lm(\hh^\dd \hh)}.
\\
\frac{E_b}{N_0}_{\min} = \min_{\substack{\K_x \succeq \0 \\ \tr(\K_x) \le 1}}\frac{\frac{\theta T P}{mN_0}}{-\log_e \E\left\{e^{-\frac{\theta T P}{mN_0} \frac{1}{\log_e 2} \, \tr(\hh \K_x \hh^\dd)}\right\}} \ge \min_{\substack{\K_x \succeq \0 \\ \tr(\K_x) \le 1}} \frac{\log_e 2} {\tr(\hh \K_x \hh^\dd)} = \frac{\log_e 2} {\lm(\E\{\hh^\dd \hh\})}
\end{gather}
Note that the right-hand sides of the above inequalities are the minimum bit energy expressions in the low-power regime and also the wideband regime with rich multipath fading due to the equivalence of the two. From this, we immediately conclude that the sparse multipath fading with bounded number of resolvable paths (or equivalently bounded number of subchannels) induces additional energy requirements in the presence of QoS constraints.
\end{rem}

\begin{rem}
Recall from the result of Theorem \ref{theo:derivsunknown} that when the transmitter has only statistical knowledge of the channel, the optimal transmission strategy in the low-power regime (and also in the wideband regime with rich multipath fading) is to transmit the information in the maximal-eigenvalue eigenspace of $\E\{\hh^\dd \hh\}$. On the other hand, we note from Theorem \ref{theo:minbitenergywb} that this is not necessarily the optimal transmission technique in the wideband regime with sparse fading. The optimal input covariance is the one that minimizes (\ref{eq:minbitenergywbKx}). Note further that for small $\frac{\theta T P}{mN_0}$, we have the following first-order Taylor series expansion of the denominator of (\ref{eq:minbitenergywbKx}):
\begin{gather}\label{eq:expansion}
-\log_e \E\left\{e^{-\frac{\theta T P}{mN_0} \frac{1}{\log_e 2} \, \tr(\hh \K_x \hh^\dd)}\right\} = \frac{\theta T P}{mN_0} \frac{1}{\log_e 2} \, \tr(\hh \K_x \hh^\dd) + o\left(\frac{\theta T P}{mN_0}\right).
\end{gather}
Hence, when $\theta$ or $P$ is small or $m$ is large, the input covariance that is optimal to the first order is the one that maximizes $\tr(\hh \K_x \hh^\dd)$, i.e., in this case, transmission in the maximal-eigenvalue eigenspace of $\E\{\hh^\dd \hh\}$ is optimal as in the low-power regime.
\end{rem}

Theorem \ref{theo:minbitenergywb} holds for the case in which the number of resolvable multipath components remains bounded. Another scenario in sparse multipath fading is the one in which the number of resolvable paths increases with bandwidth but only \emph{sublinearly}. In this case, both $B_c$ and $m$ increase without bound as $B \to \infty$ due to the sublinear growth of $B_c$. Therefore, the minimum bit energy results can be obtained by letting $m \to \infty$ in the results of Theorem \ref{theo:minbitenergywb}.

\begin{theo} \label{theo:minbitenergywblargem}
Assume a sparse multipath fading scenario in which the number of independent resolvable paths increase sublinearly with bandwidth. In this case, the minimum bit energy for a given input covariance matrix is given by
\begin{gather}
\frac{E_b}{N_0}_{\min} = \frac{\log_e2}{\tr(\hh \K_x \hh^\dd)}.
\end{gather}
When the transmitter perfectly knows the channel matrix $\hh$ and when it knows only $\E\{\hh^\dd \hh\}$, the minimum bit energies are
\begin{gather}
\frac{E_b}{N_0}_{\min} = \frac{\log_e2}{\lm(\hh^\dd \hh)}
\quad \text{and} \quad
\frac{E_b}{N_0}_{\min} = \frac{\log_e 2} {\lm(\E\{\hh^\dd \hh\})}, \label{eq:minbitenergywbknownunknownlargem}
\end{gather}
respectively.
\end{theo}

\emph{Proof:} As mentioned above, proof follows by finding the limiting values of the minimum bit energy expressions in Theorem \ref{theo:minbitenergywb} as $m \to \infty$. For the case of fixed covariance matrix $\K_x$, we have
\begin{align}
\frac{E_b}{N_0}_{\min} &= \lim_{m \to \infty} \frac{\frac{\theta T P}{mN_0}}{-\log_e \E\left\{e^{-\frac{\theta T P}{mN_0} \frac{1}{\log_e 2} \, \tr(\hh \K_x \hh^\dd)}\right\}}
\\
&= \lim_{m \to \infty} \frac{\frac{\theta T P}{mN_0}}{\frac{\theta T P}{mN_0} \frac{1}{\log_e 2} \, \tr(\hh \K_x \hh^\dd) + o\left(\frac{\theta T P}{mN_0}\right)} \label{eq:ebnominwblargem1}
\\
&= \lim_{m \to \infty} \frac{1}{ \frac{1}{\log_e 2} \, \tr(\hh \K_x \hh^\dd) + \frac{o\left(\frac{\theta T P}{mN_0}\right)}{\frac{\theta T P}{mN_0}}} \label{eq:ebnominwblargem2}
\\
&= \frac{1}{ \frac{1}{\log_e 2} \, \tr(\hh \K_x \hh^\dd) + \lim_{m \to \infty}  \frac{o\left(\frac{\theta T P}{mN_0}\right)}{\frac{\theta T P}{mN_0}}} \label{eq:ebnominwblargem3}
\\
&= \frac{\log_e 2}{\tr(\hh \K_x \hh^\dd)}. \label{eq:ebnominwblargem4}
\end{align}
(\ref{eq:ebnominwblargem1}) is obtained by using the first-order Taylor expansion in (\ref{eq:expansion}). (\ref{eq:ebnominwblargem2}) follows by dividing the numerator and denominator by $\frac{\theta TP}{mN_0}$. Finally, (\ref{eq:ebnominwblargem4}) is obtained immediately from the definition that $\lim_{x \to 0} \frac{o(x)}{x} = 0$. The expressions in (\ref{eq:minbitenergywbknownunknownlargem}) are determined as in the proofs of Theorems \ref{theo:derivsknown} and \ref{theo:derivsunknown} by choosing the input covariance matrix as $\K_x = \bu \bu^\dd$ where $\bu$ is the eigenvector that corresponds to the maximum eigenvalue of $\hh^\dd \hh$ (when $\hh$ is perfectly known at the transmitter) or of $\E\{\hh^\dd \hh\}$ (when only $\E\{\hh^\dd \hh\}$ is known at the transmitter). \hfill $\blacksquare$

\begin{rem}
Theorem \ref{theo:minbitenergywblargem} shows that as long as  the number of subchannels $m$ grows without bound in the wideband regime, the minimum bit energy requirements are the same as those in the low-power regime and wideband regime with rich multipath fading in which $m$ increases linearly with bandwidth. Note that since each subchannel experiences independent fading, $m$ can be seen as a measure of the degrees of freedom in the system. Therefore, if $m$ is bounded, the degrees of freedom is also bounded and that results in increased energy requirements as discussed in Remark \ref{rem:increasedenergy}. On the other hand, if the degrees of freedom increase with bandwidth, we have the same minimum bit energy values even though the increase is sublinear. However, for this case, we will observe in the numerical results in Section \ref{sec:numerical} that approaching the minimum bit energy is very slow and demanding in bandwidth due to zero wideband slope.
\end{rem}

\begin{rem}
Note that having $m \to \infty$ for fixed $\theta > 0$ in the minimum bit energy expressions in (\ref{eq:minbitenergywbKx})--(\ref{eq:minbitenergywbunknown}) is the same as letting $\theta \to 0$ for fixed $m$. Hence, even if $m$ is bounded, the minimum bit energies given in Theorem \ref{theo:minbitenergywblargem} are attained when $\theta = 0$. This indicates that multipath sparsity does not affect the performance in the absence of QoS constraints.
\end{rem}

\section{The Impact of QoS Constraints in the High-SNR Regime} \label{sec:highSNR}

In this section, we consider a single flat-fading channel and analyze how QoS limitations affect the performance in the high-$\tsnr$ regime. In contrast to the previous sections where general models are used, we here consider a specific fading scenario in which the components of $\hh$ are independent and identically distributed (i.i.d.) Gaussian random variables with zero mean and unit variance. Moreover, we assume that the input covariance matrix is $\K_x = \frac{1}{n_T} \I$. Note that this covariance matrix is optimal in the sense of achieving the ergodic Shannon capacity when $\hh$ has the above distribution and the transmitter does not know the realizations of $\hh$ \cite{telatar-mimo}.

Now, for the considered channel and input models, the effective rate is given by
\begin{gather}
R_{E, \text{id}}(\tsnr) =  -\frac{1}{\theta TB}\log_e\mathbb{E}\left\{\exp\left(-\theta
T B \log_2 \det \left(\I + \frac{n_R}{n_T} \tsnr \hh  \hh^\dd\right)\right)\right\} \,\, \text{bits/s/Hz}.
\end{gather}
Note that in the above formulation, we have not normalized the effective rate expression with the number of receive antennas $n_R$, and we have chosen a slightly different font from before and use the notation $R_{E,\id}$ to denote this unnormalized effective rate.

As also pointed before, the effective capacity and effective rate expressions are proportional to the logarithm of the moment generating functions of instantaneous transmission rates. For the channel and input models considered in this section, Wang and Giannakis in \cite[Theorem 1]{wang} provided an expression for the moment generating function of instantaneous mutual information. Applying this result to our setting, we obtain
\begin{align}
\E\left\{\exp\left(-\theta
T B \log_2 \det \left(\I + \frac{n_R}{n_T} \tsnr \hh  \hh^\dd\right)\right)\right\} = \frac{\det(\G(\theta,\tsnr))}{\prod_{i =1}^k \Gamma(d + i)}
\end{align}
where $\Gamma(\cdot)$ is the Gamma function, $k = \min(n_R, n_T)$, and $d = \max(n_R, n_T) - \min(n_R, n_T)$. Moreover, $\G$ is a $k \times k$ Hankel matrix whose $(i,j)^{\text{th}}$ component is
\begin{align}
g_{i,j} &= \int_0^\infty \left(1 + \frac{n_R}{n_T} \, \tsnr\, z \right)^{-\theta T B \log_2 e} z^{i + j +d} \, e^{-z} \, dz \qquad i,j = 0,1,\ldots,k-1.\label{eq:gij0}
\end{align}
Therefore, we have
\begin{gather} \label{eq:effratehighSNR}
R_{E, \text{id}}(\tsnr) =  -\frac{1}{\theta TB}\log_e \left( \frac{\det(\G(\theta,\tsnr))}{\prod_{i =1}^k \Gamma(d + i)}\right).
\end{gather}
In order to quantify the impact of the QoS constraints on the performance in the high-$\tsnr$ regime,
we consider two measures, $\si$ and $\li$, which are defined as
\begin{gather}
\si = \lim_{\tsnr \to \infty} \frac{\Ri(\tsnr)}{\log_2 \tsnr}
\intertext{and}
\li = \lim_{\tsnr \to \infty} \left( \log_2 \tsnr - \frac{\Ri(\tsnr)}{\si}\right).
\end{gather}
Note that while $\si$ denotes the high-$\tsnr$ slope in bits/s/Hz/(3dB), $\li$ represents the power offset with respect to a reference channel having the same high-$\tsnr$ slope but with unfaded and orthogonal dimensions \cite{lozano-highsnr}. With these quantities, the effective rate is approximated at high $\tsnr$s as
\begin{gather}
\Ri = \si(\log_2 \tsnr - \li) + o(1).
\end{gather}
The above high-$\tsnr$ approximation was first introduced and used in \cite{shamai-verdu} in the study of code-division multiple access systems with random spreading, and was later employed in \cite{lozano-highsnr} in the study of ergodic Shannon capacity of multiple-antenna systems. Here, we apply this approximation to the multiple-antenna systems operating under statistical queueing constraints. The next result identifies the values of $\si$ and $\li$ for a subset of values of the QoS exponent $\theta$.
\begin{theo} \label{theo:highSNR}
Assume that the components of channel matrix $\hh$ are independent and identically distributed (i.i.d.) Gaussian random variables with zero mean and unit variance. If the QoS exponent satisfies
\begin{gather} \label{eq:thetaconstraint}
\theta < \frac{\max(n_R,n_T) - \min(n_R,n_T) + 1}{TB \log_2e},
\end{gather}
then, we have
\begin{gather}
\si = \min(n_R, n_T),
\intertext{and}
\li = \left\{
\begin{array}{ll}
\log_2 \frac{n_T}{n_R} + \frac{1}{\theta T B n_R} \log_e E\left\{e^{-\theta T B \log_2 \det \hh \hh^\dd}\right\} & n_R \le n_T
\\
\log_2 \frac{n_T}{n_R} + \frac{1}{\theta T B n_T} \log_e E\left\{e^{-\theta T B \log_2 \det \hh^\dd \hh}\right\} & n_R > n_T
\end{array}\right.. \label{eq:litheta}
\end{gather}
\end{theo}

\emph{Proof}: Note that the components of the Hankel matrix $\G$, which appears in the effective rate expression in (\ref{eq:effratehighSNR}), can be written as
\begin{gather}\label{eq:gij}
g_{i,j} = \tsnr^{-\theta T B \log_2 e} \int_0^\infty \left(\frac{1}{\tsnr} + \frac{n_R}{n_T} \, z \right)^{-\theta T B \log_2 e} z^{i + j +d} \, e^{-z} \, dz \qquad i,j = 0,1,\ldots,k-1
\end{gather}
where $k = \min(n_R, n_T)$.
As $\tsnr \to \infty$, the integral in the above expression goes to a nonzero and finite value if $-\theta TB \log_2 e + i + j + d > -1$ since $0 < \int_0^\infty z^a e^{-z} dz < \infty$  for $a>-1$ and $\int_0^\infty z^a e^{-z} dz = \infty$ for $a \le -1$. Note that this condition is satisfied for all $i,j = 0,1,\ldots,k-1$ by our assumption in (\ref{eq:thetaconstraint}). Now, we can immediately see that $g_{i,j}$ for all $i,j$ scales as $\tsnr^{-\theta T B \log_2 e}$ as $\tsnr \to \infty$. Therefore, the determinant of $\G$ scales as $\tsnr^{-k \theta T B \log_2 e}$. This lets us conclude that
\begin{align}
R_{E, \text{id}}(\tsnr) =  -\frac{1}{\theta TB}\log_e \left( \frac{\det(\G(\theta,\tsnr))}{\prod_{i =1}^k \Gamma(d + i)}\right) &= -\frac{1}{\theta TB}\log_e\left( \tsnr^{-k \theta T B \log_2 e} \right) + O(1)
\\
&= k (\log_2e) \log_e \tsnr + O(1)
\\
&=k \log_2 \tsnr + O(1)
\\
&= \min(n_R,n_T) \log_2 \tsnr + O(1),
\end{align}
establishing that $\si = \min(n_R,n_T)$ for the values of $\theta$ specified in the theorem. Above, $O(1)$ denotes the terms that approach a finite constant as $\tsnr \to \infty$.

Next, we consider the power offset $\li$. Assume that $n_R \le n_T$. Under this assumption, we have
\begin{align}
\li &= \lim_{\tsnr \to \infty} \left( \log_2 \tsnr - \frac{\Ri(\tsnr)}{\si}\right)
\\
&= \lim_{\tsnr \to \infty} \left( \log_2 \tsnr - \frac{\Ri(\tsnr)}{n_R}\right)
\\
&= \lim_{\tsnr \to \infty} \left( \log_2 \tsnr + \frac{\frac{1}{\theta TB}\log_e\mathbb{E}\left\{e^{-\theta
    T B \log_2 \det \left(\I + \frac{n_R}{n_T} \tsnr \, \hh  \hh^\dd\right)}\right\}}{n_R}\right)
\\
&= \lim_{\tsnr \to \infty} \left( \log_2 \tsnr + \frac{\frac{1}{\theta TB}\log_e\mathbb{E}\left\{e^{-\theta
T B n_R \log_2 \tsnr -\theta
T B \log_2 \det \left(\frac{1}{\tsnr}\I + \frac{n_R}{n_T} \hh  \hh^\dd\right)}\right\}}{n_R}\right) \label{eq:li1}
\end{align}
\begin{align}
&= \lim_{\tsnr \to \infty} \left( \log_2 \tsnr + \frac{ -n_R \log_2 \tsnr + \frac{1}{\theta TB}\log_e\mathbb{E}\left\{e^{-\theta
T B \log_2 \det \left(\frac{1}{\tsnr}\I + \frac{n_R}{n_T} \hh  \hh^\dd\right)}\right\}}{n_R}\right)
\\
&= \lim_{\tsnr \to \infty} \frac{1}{\theta TBn_R}\log_e\mathbb{E}\left\{e^{-\theta
T B \log_2 \det \left(\frac{1}{\tsnr}\I + \frac{n_R}{n_T} \hh  \hh^\dd\right)}\right\}
\\
&= \frac{1}{\theta TBn_R}\log_e\mathbb{E}\left\{e^{-\theta
T B \log_2 \det \left(\frac{n_R}{n_T} \hh  \hh^\dd\right)}\right\}
\\
&= \log_2 \frac{n_T}{n_R} + \frac{1}{\theta TBn_R}\log_e\mathbb{E}\left\{e^{-\theta
T B \log_2 \det \hh  \hh^\dd}\right\}.
\end{align}
Above, while (\ref{eq:li1}) is obtained by noting that $$\theta
T B \log_2 \det \left(\I + \frac{n_R}{n_T} \tsnr \hh  \hh^\dd\right) = \theta
T B \log_2 \tsnr^{n_R} + \theta
T B \log_2 \det \left(\frac{1}{\tsnr}\I + \frac{n_R}{n_T} \hh  \hh^\dd\right),$$
the remaining steps follow through straightforward algebraic operations. The result for the case in which $n_T < n_R$ can be readily proved by applying the above procedure to
\begin{gather}
\li = \lim_{\tsnr \to \infty} \left( \log_2 \tsnr + \frac{\frac{1}{\theta TB}\log_e\mathbb{E}\left\{e^{-\theta
T B \log_2 \det \left(\I + \frac{n_R}{n_T} \tsnr\,\hh^\dd  \hh\right)}\right\}}{n_T}\right).
\end{gather}
\hfill $\blacksquare$

\begin{rem}
When ergodic Shannon rate (or equivalently effective rate with $\theta = 0$) is considered, it is well-known that the high-$\tsnr$ slope for the i.i.d. Rayleigh fading channel is $\si = \min(n_R,n_T)$. The above result shows that the high-$\tsnr$ slope does not get affected by the queueing constraints when $\theta < \frac{\max(n_R,n_T) - \min(n_R,n_T) + 1}{TB \log_2e}$.
\end{rem}

\begin{rem}
For the case of $\theta = 0$, it is shown in \cite[Appendix B]{lozano-highsnr} that the power offset in the i.i.d. Rayleigh fading is\footnote{In \cite{lozano-highsnr}, signal-to-noise ratio is defined as $\tsnr = \frac{n_R \E\{\|\x\|^2\}}{\E\{\|\n\|^2\}}$. Due to the presence of $n_R$ in the numerator in the $\tsnr$ definition, the first term of $\li$ in \cite{lozano-highsnr} is $\log_2 n_T$ instead of $\log_2 \frac{n_T}{n_R}$.}
\begin{gather}\label{eq:litheta0}
\li = \left\{
\begin{array}{ll}
\log_2 \frac{n_T}{n_R} - \frac{1}{n_R} E\left\{\log_2 \det \hh \hh^\dd\right\} & n_R \le n_T
\\
\log_2 \frac{n_T}{n_R} - \frac{1}{n_T} E\left\{ \log_2 \det \hh^\dd \hh\right\} & n_R > n_T
\end{array}\right..
\end{gather}
By Jensen's inequality and strict concavity of the logarithm function, we have
\begin{align}
\frac{1}{\theta T B n_R}\log_e E\left\{e^{-\theta T B \log_2 \det \hh \hh^\dd}\right\} & > \frac{1}{\theta T B n_R} E\left\{ \log_e e^{-\theta T B \log_2 \det \hh \hh^\dd}\right\}
\\
&=-\frac{1}{n_R} E\left\{ \log_2 \det \hh \hh^\dd\right\}, \quad \text{for} \quad \theta > 0
\end{align}
which shows from the comparison of (\ref{eq:litheta}) and (\ref{eq:litheta0}) that the presence of queueing constraints result in higher power offset values in the high-$\tsnr$ regime.
\end{rem}

\begin{rem}
Note that by H\"{o}lder's inequality, we have
\begin{gather}
(E\{|x|^r\})^{1/r} \le (E\{|x|^s\})^{1/s} \label{eq:holderineq}
\end{gather}
for $0 < r < s$. Note further that the second term in the expression of $\li$ can be expressed as \footnote{Without loss of generality, we consider the case in which $n_R \le n_T$.}
\begin{align}
\frac{1}{\theta T B n_R} \log_e E\left\{e^{-\theta T B \log_2 \det \hh \hh^\dd}\right\} = \frac{1}{n_R} \log_e \left( E\left\{e^{-\theta T B \log_2 \det \hh \hh^\dd}\right\}\right)^{\frac{1}{\theta T B}}.
\end{align}
Application of the inequality in (\ref{eq:holderineq}) to $\left(E\left\{e^{-\theta T B \log_2 \det \hh \hh^\dd}\right\}\right)^{\frac{1}{\theta T B}}$ shows that the power offset $\li$ in a non-decreasing function of the QoS exponent $\theta$.
\end{rem}

Theorem \ref{theo:highSNR} characterizes $\si$ and $\li$ for a certain range of values of $\theta$. The next result gives a partial answer to what is expected when $\theta > \frac{\max(n_R,n_T) - \min(n_R,n_T) + 1}{TB \log_2e}$, by considering the case of single-antenna transmission and reception, i.e., $n_T = n_R = 1$.

\begin{theo} \label{theo:si-singleantenna}
In a Rayleigh fading channel with single transmit antenna and single receive antenna (i.e., $n_T = n_R = 1$), the high-$\tsnr$ slope is
\begin{gather}
\si = \frac{1}{\theta T B \log_2 e}
\end{gather}
when $\theta > \frac{1}{T B \log_2 e}$.
\end{theo}

\emph{Proof}: When we have $n_T = n_R = 1$, the effective rate expression is
\begin{align}
R_{E}(\tsnr) &=  -\frac{1}{\theta TB}\log_e \mathbb{E}\left\{e^{-\theta
T B \log_2 \left(1 + \tsnr |h|^2 \right)}\right\}
\\
&=  -\frac{1}{\theta TB}\log_e \mathbb{E}\left\{e^{\log_e \left(1 + \tsnr |h|^2 \right)^{-\theta
T B \log_2 e}}\right\}
\\
&=  -\frac{1}{\theta TB}\log_e \mathbb{E}\left\{\left(1 + \tsnr |h|^2 \right)^{-\theta
T B \log_2 e}\right\}
\\
&=  -\frac{1}{\theta TB}\log_e \int_0^\infty \left(1 + \tsnr z \right)^{-\theta
T B \log_2 e} e^{-z} \, dz \label{eq:effrate-singleantenna}
\end{align}
where (\ref{eq:effrate-singleantenna}) follows from our Rayleigh fading assumption which implies that $z = |h|^2$ has an exponential distribution. Note that this effective rate expression can also be immediately seen to be a special case of the expressions in (\ref{eq:gij0}) and (\ref{eq:effratehighSNR}). Now, we prove the result through the following steps:
\begin{align}
\si &= \lim_{\tsnr \to \infty} \frac{R_E(\tsnr)}{\log_2 \tsnr}
\\
&= \lim_{\tsnr \to \infty} \frac{-\frac{1}{\theta TB}\log_e \int_0^\infty \left(1 + \tsnr z \right)^{-\theta
T B \log_2 e} e^{-z} \, dz}{\log_2 \tsnr} \label{eq:si-singleantenna0}
\\
&= \lim_{\tsnr \to \infty} \frac{-\frac{1}{\theta TB} \log_e \left[ \frac{\tsnr}{\tsnr} \int_0^\infty \left(1 + \tsnr z \right)^{-\theta
T B \log_2 e} e^{-z} \, dz \right]}{\log_2 \tsnr} \label{eq:si-singleantenna1}
\\
&= \lim_{\tsnr \to \infty} \frac{\frac{1}{\theta TB}\log_e \tsnr -\frac{1}{\theta TB}\log_e \left[\tsnr \int_0^\infty \left(1 + \tsnr z \right)^{-\theta
T B \log_2 e} e^{-z} \, dz \right]}{\log_2 \tsnr}\label{eq:si-singleantenna2}
\\
&= \lim_{\tsnr \to \infty} \frac{\frac{1}{\theta TB}\log_e \tsnr}{\log_2 \tsnr} +\frac{-\frac{1}{\theta TB}\log_e \left[ \tsnr \int_0^\infty \left(1 + \tsnr z \right)^{-\theta
T B \log_2 e} e^{-z} \, dz \right]}{\log_2 \tsnr} \label{eq:si-singleantenna3}
\\
&= \frac{1}{\theta T B \log_2 e} + \lim_{\tsnr \to \infty} \frac{-\frac{1}{\theta TB}\log_e \left[\tsnr^{-\theta
T B \log_2 e + 1} \int_0^\infty \left(\frac{1}{\tsnr} + z \right)^{-\theta
T B \log_2 e} e^{-z} \, dz \right]}{\log_2 \tsnr} \label{eq:si-singleantenna4}
\\
&= \frac{1}{\theta T B \log_2 e} + \lim_{\tsnr \to \infty} \frac{-\frac{1}{\theta TB}\log_e \left[\tsnr^{-\theta
T B \log_2 e + 1} \, e^{\frac{1}{\ssnr}} \, \Gamma\left(-\theta
T B \log_2 e + 1, \frac{1}{\ssnr}\right)\right]}{\log_2 \tsnr} \label{eq:si-singleantenna5}
\\
&= \frac{1}{\theta T B \log_2 e} + \lim_{\tsnr \to \infty} \frac{-\frac{1}{\theta TB}\log_e \left[e^{\frac{1}{\ssnr}} \, \frac{\Gamma\left(-\theta
T B \log_2 e + 1, \frac{1}{\tisnr}\right)}{\frac{1}{\tisnr}^{-\theta
T B \log_2 e + 1}}\right]}{\log_2 \tsnr} \label{eq:si-singleantenna6}
\\
&= \frac{1}{\theta T B \log_2 e}. \label{eq:si-singleantenna7}
\end{align}
Above, (\ref{eq:si-singleantenna2}) is obtained by multiplying the integral inside the logarithm in the numerator by $\frac{\tsnr}{\tsnr}$ as shown in (\ref{eq:si-singleantenna1}). and by using the fact that the logarithm of the division is equal to the difference of the logarithms.  (\ref{eq:si-singleantenna3}) follows by separately writing the fractions. (\ref{eq:si-singleantenna4}) is obtained by evaluating the limit of the first fraction, and by expressing $ \left(1 + \tsnr z \right)^{-\theta
T B \log_2 e}$ in the second fraction as $\tsnr^{-\theta
T B \log_2 e} \left(\frac{1}{\tsnr} + z \right)^{-\theta
T B \log_2 e}$. (\ref{eq:si-singleantenna5}) follows from the fact that \cite[Equation 3.382.4]{integralbook}
\begin{gather}
\int_0^\infty \left(\frac{1}{\tsnr} + z \right)^{-\theta
T B \log_2 e} e^{-z} \, dz = e^{\frac{1}{\ssnr}} \, \Gamma\left(-\theta
T B \log_2 e, \frac{1}{\ssnr}\right)
\end{gather}
where
$\Gamma(\alpha, x)$ is the upper incomplete Gamma function. (\ref{eq:si-singleantenna6}) is obtained by rearranging the terms in the numerator of the fraction in the second term. Finally, (\ref{eq:si-singleantenna7}) follows by realizing that the limiting expression in (\ref{eq:si-singleantenna6}) is equal to zero. This is noted from the fact that as $\tsnr \to \infty$, we have
\begin{gather}
e^{\frac{1}{\ssnr}} \longrightarrow 1
\\
\frac{\Gamma\left(-\theta
T B \log_2 e + 1, \frac{1}{\tisnr}\right)}{\frac{1}{\tisnr}^{-\theta
T B \log_2 e + 1}} \longrightarrow \frac{1}{\theta T B \log_2 e -1}, \label{eq:Gammalimit}
\end{gather}
indicating that the numerator in the limiting expression in (\ref{eq:si-singleantenna6}) is approaching a finite value as $\tsnr$ increases while the denominator grows without bound. The limit in (\ref{eq:Gammalimit}) is due to the fact that
\footnote{The limit in \eqref{eq:originalgammalimit} can be obtained from the following facts: A definition of the upper incomplete Gamma function is given by  \cite[Equation 8.351.4]{integralbook} $\Gamma(\alpha,x) = x^\alpha e^{-x} \Psi(1,1+\alpha; x) = x^\alpha e^{-x} \int_0^\infty e^{-xt} (1+t)^{\alpha-1} dt$. From this definition, we can easily see that $\lim_{x \to 0} \frac{\Gamma(\alpha,x)}{x^\alpha} = \int_0^\infty (1+t)^{\alpha-1} dt = \frac{-1}{\alpha}$ for $\alpha < 0$.}
\begin{gather} \label{eq:originalgammalimit}
\frac{\Gamma(\alpha, x)}{x^\alpha} \to \frac{-1}{\alpha} \quad \text{as} \quad x \to 0
\end{gather}
when $\alpha < 0$, which is satisfied in our setting from our assumption that $\theta T B \log_2 e > 1$. \hfill $\blacksquare$

\begin{rem}
Theorem \ref{theo:si-singleantenna} shows for the single-antenna case that when $\theta > \frac{\max(n_R,n_T) - \min(n_R,n_T) + 1}{TB \log_2e} = \frac{1}{T B \log_2 e}$, the high-$\tsnr$ slope is $\si = \frac{1}{\theta T B \log_2e} < \min(n_R, n_T) = 1$, and diminishes with increasing $\theta$. Note that by Theorem \ref{theo:highSNR}, $\si = \min(n_R, n_T)=1$ when $\theta < \frac{1}{TB \log_2 e}$ in the case of single antennas at the receiver and transmitter.
\end{rem}

\begin{rem}
For the multiple-antenna case, we have the following additional discussion. An expression for the components of the Hankel matrix $\G$ is given by \cite{wang}
\begin{align}
g_{i,j} &= \int_0^\infty \left(1 + \frac{n_R}{n_T} \, \tsnr\, z \right)^{-\theta T B \log_2 e} z^{i + j +d} \, e^{-z} \, dz \qquad i,j = 0,1,\ldots,k-1
\\
&= \frac{\pi}{\Gamma\left(\theta T B \log_2 e\right) \sin\left(\pi\left(d + i + j - \theta T B \log_2 e\right)\right)} \times \nonumber
\\
&\hspace{.4cm}\Bigg[\frac{\left(\frac{n_R}{n_T} \, \tsnr \right)^{-1-d-i-j} \Gamma(1 + d + i + j)}{\Gamma\left(2 + d + i + j -\theta T B \log_2 e  \right)} \, {_1}F_1\left(1+d+i+j, 2 + d + i + j - \theta T B \log_2 e, \frac{n_T}{n_R \tsnr} \right)  \nonumber
\\
&\hspace{.7cm} - \frac{\left(\frac{n_R}{n_T} \, \tsnr \right)^{-\theta T B \log_2 e} \Gamma(\theta T B \log_2 e)}{\Gamma\left(-d - i - j + \theta T B \log_2 e  \right)} \, {_1}F_1\left(\theta T B \log_2 e, - d - i - j + \theta T B \log_2 e, \frac{n_T}{n_R \tsnr} \right) \Bigg] \label{eq:gijexpression}
\end{align}
where ${_1}F_1$ denotes the confluent hypergeometric function and has the following series expansion \cite{integralbook}
\begin{align}
{_1}F_1(a,b,z) =\sum_{i = 0}^\infty \frac{(a)_i z^i}{(b)_i i!} = 1 + \frac{a}{b}\frac{z}{1!} + \frac{a(a+1)}{b(b+1)} \frac{z^2}{2!} + \frac{a(a+1)(a+2)}{b(b+1)(b+2)} \frac{z^3}{3!} + \ldots
\end{align}
Note that the expression in (\ref{eq:gijexpression}) is valid when $\theta T B \log_2 e \neq \pm (d + i + j)$ for all $i,j$ because of the presence of the sinousoid in the denominator of the first term and the fact that $\Gamma(x) = \infty$ or $-\infty$ when $x$ is a negative integer. Under this restriction, we can see (by also noting that ${_1}F_1(a,b,0) = 1$) that the first term inside the square brackets in (\ref{eq:gijexpression}) scales as $\tsnr^{-1-d-i-j}$ while the second term scales as $\tsnr^{-\theta T B \log_2 e}$ as $\tsnr \to \infty$. Note that $d = \max(n_R, n_T) - \min(n_R, n_T)$ and $i,j = 0,1,\ldots, \min(n_R,n_T)-1$. Therefore, when
\begin{gather}
\theta T B \log_2 e > 1 + d + 2(\min(n_R,n_T)-1) = \max(n_R, n_T) + \min(n_R,n_T) -1,
\end{gather}
the first terms with $\tsnr^{-1-d-i-j}$ will dictate the rate at which $g_{i,j}$'s approach zero for all $i,j$. Hence, we have
\begin{gather} \label{eq:gijhighsnrapprox}
g_{i,j} \sim \frac{\pi}{\Gamma\left(\theta T B \log_2 e\right) \sin\left(\pi\left(d + i + j - \theta T B \log_2 e\right)\right)} \frac{\left(\frac{n_R}{n_T} \, \tsnr \right)^{-1-d-i-j} \Gamma(1 + d + i + j)}{\Gamma\left(2 + d + i + j -\theta T B \log_2 e  \right)}
\end{gather}
as $\tsnr \to \infty$. Note that the matrix $\widetilde{\G}$, whose components $\tilde{g}_{i,j}$ are equal to the right-hand side of (\ref{eq:gijhighsnrapprox}), is still a Hankel matrix as the components depend on the indexes only through $(i + j)$. If the determinant is nonzero, it can be easily verified that the determinant of $\widetilde{\G}$ scales as
\begin{gather}
\det(\widetilde{\G}) \sim \tsnr^{-\sum_{i = 1}^{\min(n_R,n_T)}(2i-1)} = \tsnr^{(\min(n_R,n_T))^2}.
\end{gather}
For instance,
\begin{align}
\det(\widetilde{\G})= \det\left(\left[
\begin{array}{ccc}
  a \tsnr^{-1} & b \tsnr^{-2} & c \tsnr^{-3} \\
  b \tsnr^{-2} & c \tsnr^{-3} & d \tsnr^{-4} \\
  c \tsnr^{-3} & d \tsnr^{-4} & e \tsnr^{-5}
\end{array}\right]\right) \sim \tsnr^{-(1+3+5)} = \tsnr^{-9}
\end{align}
for large $\tsnr$ as long as the constant $a,b,c,d,$ and $e$ are such that $\det(\widetilde{\G})$ is nonzero. Finally, we have under the aforementioned conditions that
\begin{gather}
R_{E, \text{id}}(\tsnr) \sim  -\frac{1}{\theta TB}\log_e \left( \frac{\det(\widetilde{\G}(\theta,\tsnr))}{\prod_{i =1}^k \Gamma(d + i)}\right) \sim \frac{(\min(n_R,n_T))^2}{\theta T B \log_2 e} \log_2 \tsnr,
\end{gather}
indicating that
\begin{gather}
\si = \frac{(\min(n_R,n_T))^2}{\theta T B \log_2 e}
\end{gather}
when $\theta T B \log_2 e >  \max(n_R, n_T) + \min(n_R,n_T) -1$. Note that under this condition on $\theta$, $\si = \frac{\min(n_R,n_T)^2}{\theta T B \log_2 e} < \min(n_R,n_T)$. Note also that the above conclusion reduces to the result of Theorem \ref{theo:si-singleantenna} when $n_R = n_T  = 1$.
\end{rem}


\section{Numerical Results} \label{sec:numerical}

In this section, we numerically illustrate the analytical results obtained in the previous sections. In order to treat the low-$\tsnr$ and high-$\tsnr$ regimes jointly, we consider the i.i.d. Rayleigh fading channel in which the components of the channel matrix $\hh$ are i.i.d. zero-mean, unit-variance, circularly symmetric Gaussian random variables. We further assume that the input covariance matrix is $\K_x = \frac{1}{n_T}\I$, and the effective rate is given by
\begin{gather}
R_{E, \text{id}}(\tsnr) =  -\frac{1}{\theta TB}\log_e\mathbb{E}\left\{\exp\left(-\theta
T B \log_2 \det \left(\I + \frac{n_R}{n_T} \tsnr \hh  \hh^\dd\right)\right)\right\} \,\, \text{bits/s/Hz}.
\end{gather}
Under these assumptions, we can easily compute the effective rate by using the formulation in (\ref{eq:effratehighSNR}) and performing integral computations. We note that the computations of the effective rate in the correlated fading case can be done using the expressions of the moment generating function of the mutual information of correlated MIMO Gaussian fading channels provided in \cite{mario}. Summary of such non-asymptotic results, along with asymptotic spectrum theorems, on random matrices is presented in \cite{tulino-verdu}.

\begin{figure}
\begin{center}
\includegraphics[width=\figsize\textwidth]{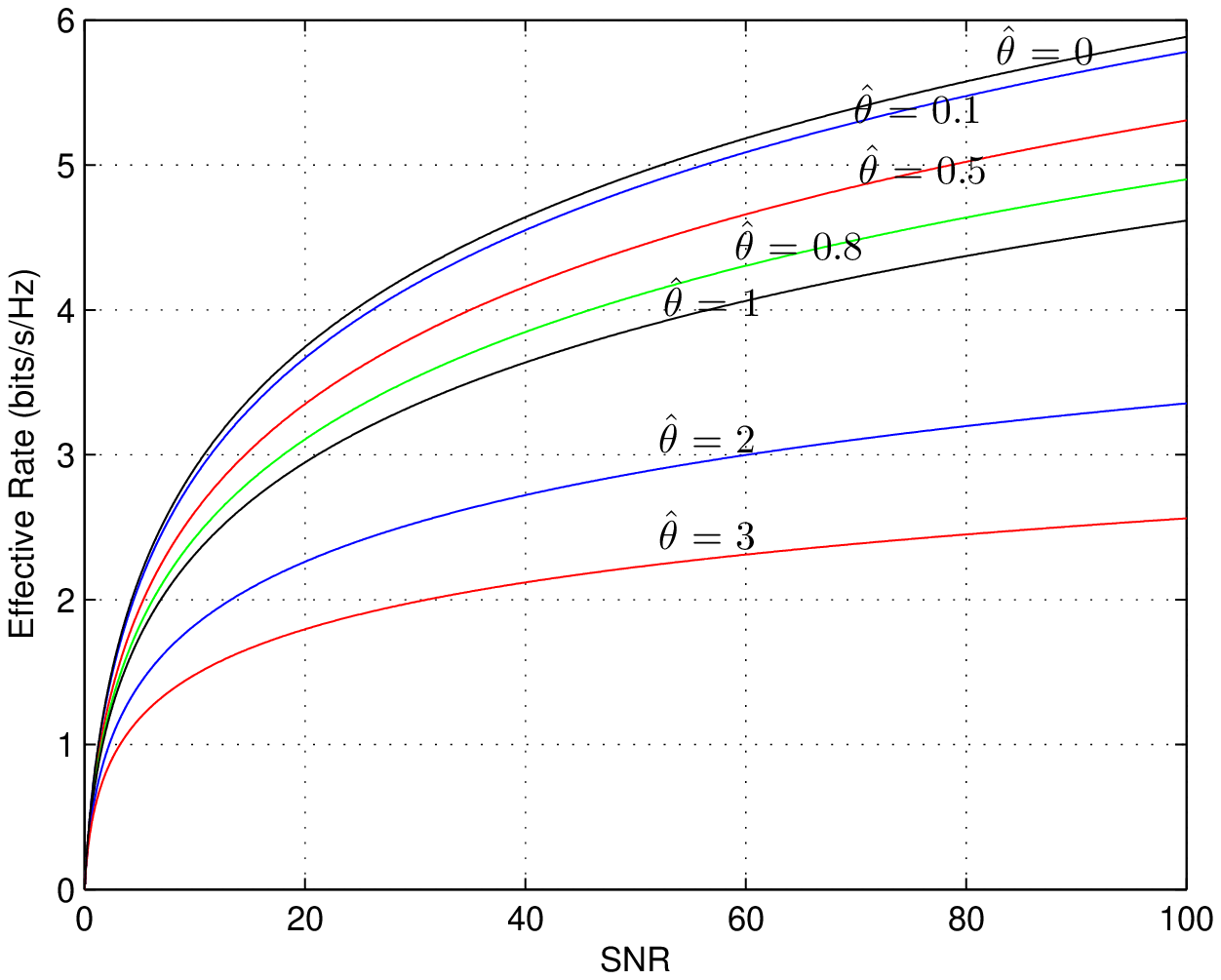}
\caption{Effective rate $\R_E$ vs. $\tsnr $ in the single-antenna case (i.e., when $n_R = n_T = 1$) for different values of $\hat{\theta} = \theta T B \log_2e$.}\label{fig:nR1nT1snrtheta0}
\end{center}
\end{figure}

Figure \ref{fig:nR1nT1snrtheta0} plots the effective rate $\Ri$ as a function of $\tsnr$ in the single-antenna case ($n_R = n_T = 1$) for different values of $\hat{\theta} = \theta T B \log_2e$. It is assumed that $T = 1\text{ ms} = 10^{-3} \text{ s}$ and $B = 100$kHz $= 10^5$Hz. Note that when $\hat{\theta} = 0$ or equivalently $\theta = 0$, there are no statistical queueing constraints and the effective capacity is equal to the ergodic Shannon capacity. In Fig. \ref{fig:nR1nT1snrtheta0}, we observe that the effective rate in general diminishes with increasingly more strict queueing constraints (or equivalently higher $\theta$ values). As expected, under more strict buffer constraints, lower arrival rates are supported, and as a result, lower departure rates are seen. On the other hand, as predicted by the low-SNR results of Section \ref{sec:lowpower}, all rate curves have the same slope at $\tsnr = 0$. Note that this slope is the one achieved in the absence of QoS constraints (i.e., when $\theta = 0$). Therefore, the impact of queueing constraints on the performance lessens at low $\tsnr$ values. An intuitive explanation of this observation is that as power decreases, arrival rates that can be supported by the system diminishes as well, which in turn decreases the effect of buffer violation constraints. Note also that as discussed in Section \ref{sec:wideband}, results similar to those in the low-power regime are obtained in the wideband regime if the channel experiences rich multipath fading. Therefore, another interpretation of the above observation is that QoS constraints have less impact on the performance as the bandwidth increases in rich multipath environments. This is due to the fact that the number of noninteracting subchannels and hence the number of degrees of freedom increases with increasing bandwidth, and the system has increasingly higher diversity to combat with buffer constraints.

Fig. \ref{fig:nR1nT1snrtheta0} confirms the analytical high-$\tsnr$ results as well. As predicted by Theorem \ref{theo:highSNR}, the high-$\tsnr$ slope is the same as that achieved in the absence of QoS constraints as long as $\hat{\theta} = \theta T B \log_2e < 1$. On the other hand, as proved in Theorem \ref{theo:si-singleantenna}, high-$\tsnr$ slope is strictly less than 1 when $\hat{\theta} > 1$. The difference in the rates of increase at high $\tsnr$s is clearly seen in Fig. \ref{fig:nR1nT1snrtheta0}.

\begin{figure}
\begin{center}
\includegraphics[width=\figsize\textwidth]{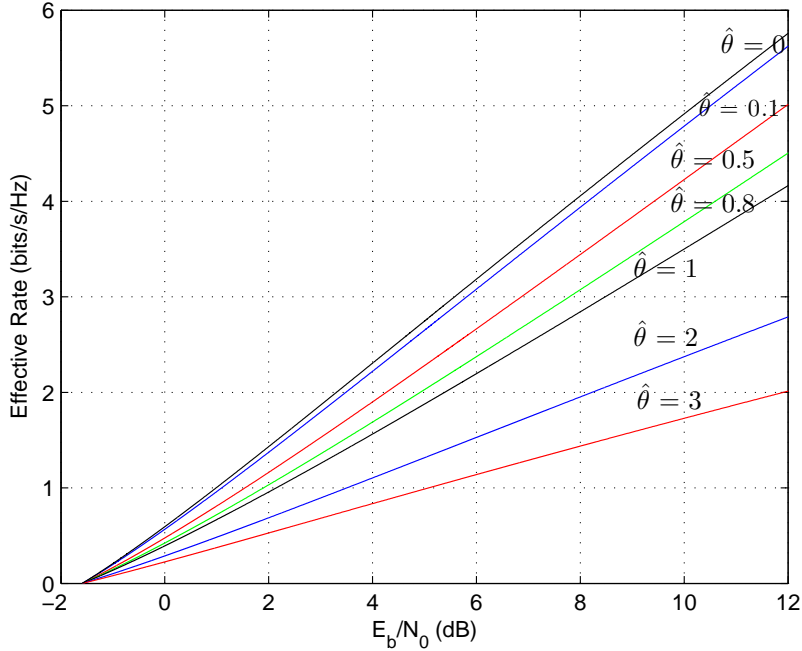}
\caption{Effective rate $\R_E$ vs. bit energy $\frac{E_b}{N_0}$ in the single-antenna case (i.e., when $n_R = n_T = 1$) for different values of $\hat{\theta} = \theta T B \log_2e$.}\label{fig:nR1nT1ebnotheta0}
\end{center}
\end{figure}

In Fig. \ref{fig:nR1nT1ebnotheta0}, we plot the effective rate as a function of the bit energy in the single-antenna case. Confirming the discussion in Section \ref{subsec:energy}, we immediately note that the minimum bit energy for all values of $\theta$ is $-1.59$ dB, which is the fundamental limit in the absence of QoS limitations. This is a consequence of the fact that the effective rate curves as a function of $\tsnr$ have the same slope at zero $\tsnr$. However, since the second derivatives of the effective rate at $\tsnr = 0$ decreases with increasing $\theta$, we observe in Fig. \ref{fig:nR1nT1ebnotheta0} that we have smaller wideband slopes, $\so$, for larger values of $\theta$. Similarly as in Fig. \ref{fig:nR1nT1snrtheta0}, we observe smaller high-$\tsnr$ slopes, $\si$, when $\hat{\theta} > 1$.

\begin{figure}
\begin{center}
\includegraphics[width=\figsize\textwidth]{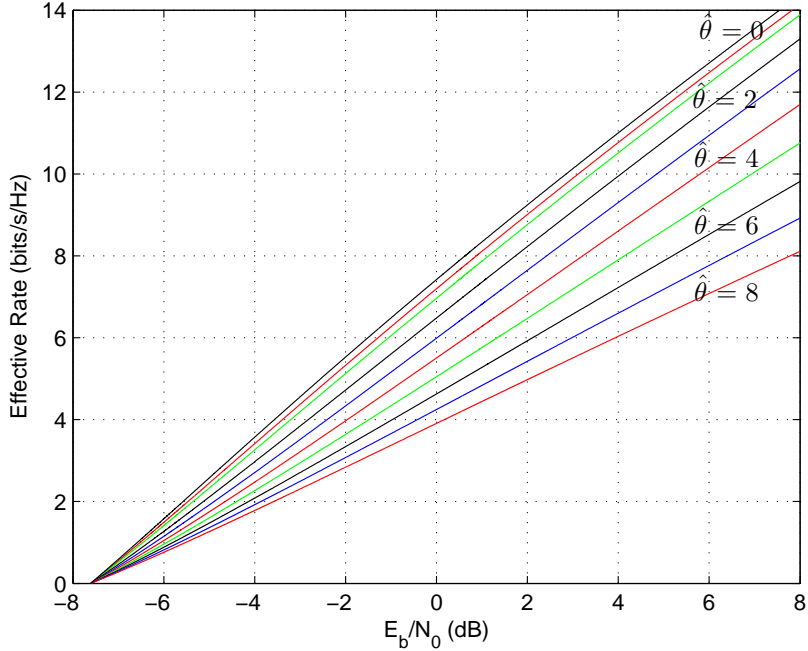}
\caption{Effective rate $\R_E$ vs. bit energy $\frac{E_b}{N_0}$ for $\hat{\theta} = \theta T B \log_2e = 0, 0.5, 1, 2, 3, 4, 5, 6, 7, 8$ when $n_R = 2$ and $n_T = 5$.}\label{fig:nR2nT5ebnotheta0}
\end{center}
\end{figure}

\begin{figure}
\begin{center}
\includegraphics[width=\figsize\textwidth]{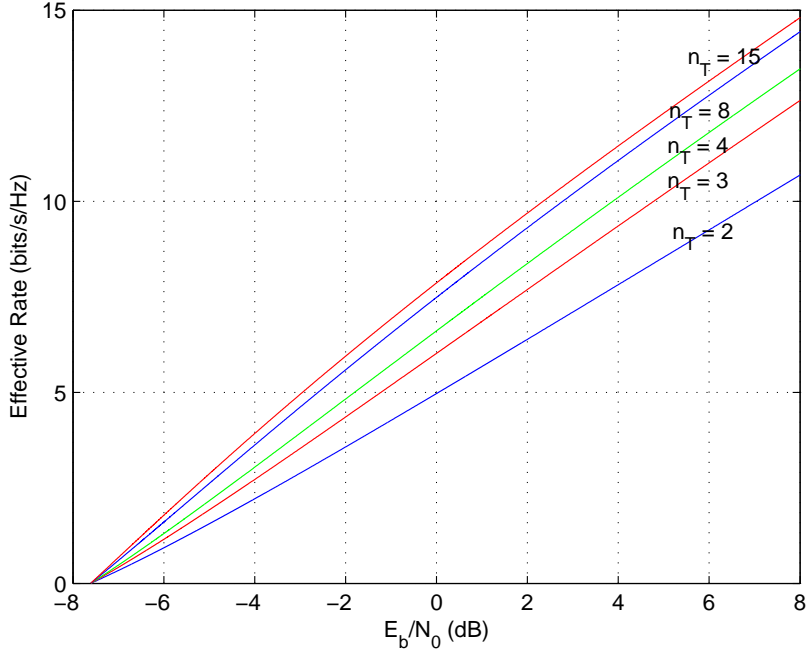}
\caption{Effective rate $\R_E$ vs. bit energy $\frac{E_b}{N_0}$ for $n_T = 2, 3, 4, 8, 15$ when $n_R = 2$ and $\hat{\theta} = \theta TB \log_2e = 1$.}\label{fig:nR2theta1ebno}
\end{center}
\end{figure}

In Fig. \ref{fig:nR2nT5ebnotheta0}, effective rate vs. bit energy curves are plotted under the assumption that the number of receive antennas is $n_R = 2$ and the number of transmit antennas is $n_T =5$. We still assume that $T = 1$ ms and $B = 100$kHz. In the figure, the curves from the top to the bottom are for $\hat{\theta} = 0, 0.5, 1, 2, 3, 4, 5, 6, 7, 8 $ in this order \footnote{Note that when $\theta = 0$, effective capacity becomes equal to the ergodic Shannon capacity. For this case, rate is computed using the formulation provided in \cite[Theorem 2]{telatar}.}. We again immediately note that the same minimum bit energy is attained for all values of $\hat{\theta}$ while the wideband slopes $\so$ are smaller for larger values of the QoS exponent. In this case, the minimum bit energy is $\frac{E_b}{N_0}_{\min} = 10\log_{10}\left(\frac{\log_e2}{n_R^2}\right) = -7.61$ dB \footnote{As opposed to (\ref{eq:minenergy-iidRayleigh}) where $\frac{E_b}{N_0}_{\min} = \frac{\log_e 2}{n_R}$, we have $\frac{E_b}{N_0}_{\min} = \frac{\log_e 2}{n_R^2}$ in the figure since we plot the effective rate in bits/s/Hz without normalization with the number of receive antennas.}. At high $\tsnr$ levels, we observe that, as shown in Theorem \ref{theo:highSNR}, when $\hat{\theta} = \theta T B \log_2e < \max(n_R,n_T) - \min(n_R,n_T) + 1 = 4$, $\si$ is the same as that achieved when $\hat{\theta} = 0$ (i.e., when $\theta = 0$). For $\hat{\theta} > 4$, we note the gradual decrease in the high-$\tsnr$ slope.

When we compare Figs. \ref{fig:nR1nT1ebnotheta0} and \ref{fig:nR2nT5ebnotheta0}, we see that the rate curves are much closer to each other in Fig. \ref{fig:nR2nT5ebnotheta0}, indicating the resilience provided by spatial diversity against queueing constraints. This is further illustrated in Fig. \ref{fig:nR2theta1ebno}, where effective rate vs. bit energy curves are plotted for different number of transmit antennas when $n_R = 2$ and $\hat{\theta} = 1$. In this figure, we observe that the wideband slope $\so$ increases with increasing number of transmit antennas for a given QoS exponent $\theta$. Moreover, we note that improvements are provided at all $\tsnr$ levels when the number of antennas is increased in the system, again pointing to the benefits of spatial diversity.

\begin{figure}
\begin{center}
\includegraphics[width=\figsize\textwidth]{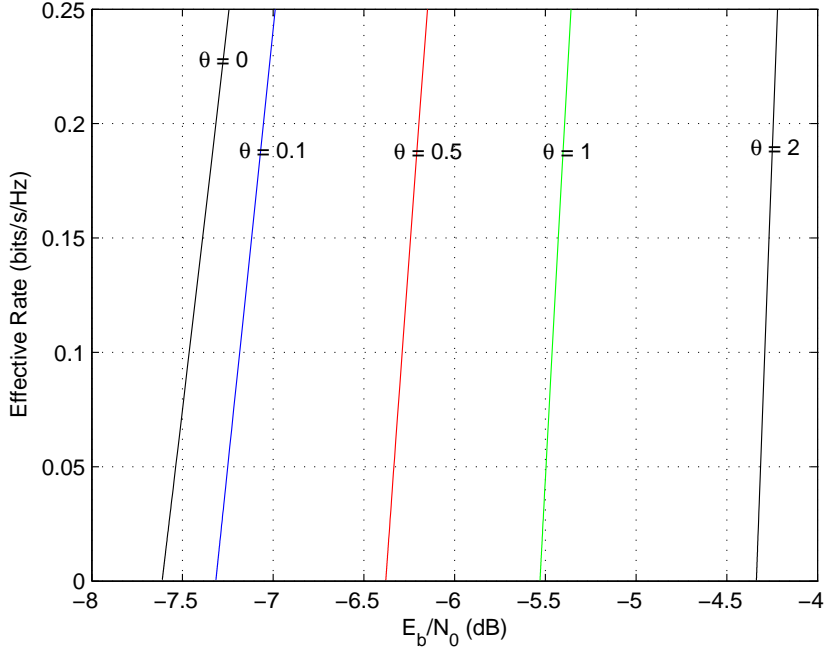}
\caption{Effective rate $\R_E$ vs. bit energy $\frac{E_b}{N_0}$ for $\theta = 0, 0.1, 0.5, 1, 2$ in sparse wideband fading channels. $n_R = n_T = 2$. The number of subchannels is $m = 5$. The coherence bandwidth $B_c$ increases with increasing bandwidth.}\label{fig:nR2nT2ebno-wideband}
\end{center}
\end{figure}

\begin{figure}
\begin{center}
\includegraphics[width=\figsize\textwidth]{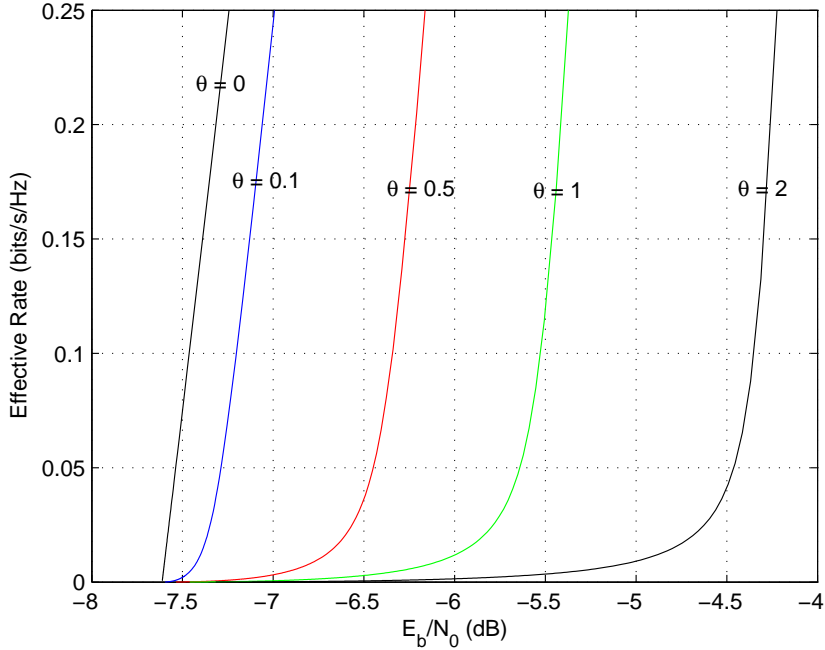}
\caption{Effective rate $\R_E$ vs. bit energy $\frac{E_b}{N_0}$ for $\theta = 0, 0.1, 0.5, 1, 2$ in sparse wideband fading channels. $n_R = n_T = 2$. Both the coherence bandwidth $B_c$ and the number of subchannels $m$ increase with increasing bandwidth. }\label{fig:nR2nT2ebno-wideband_m}
\end{center}
\end{figure}

Heretofore, the discussions on the low-$\tsnr$ regime apply to the cases in which the transmit power is small or the bandwidth is large but in a rich multipath fading setting. In Section \ref{sec:wideband}, we have remarked that sparse multipath fading has considerable impact on the performance in the wideband regime. In order to numerically illustrate these results, we provide Figs. \ref{fig:nR2nT2ebno-wideband} and \ref{fig:nR2nT2ebno-wideband_m}. In Fig. \ref{fig:nR2nT2ebno-wideband}, effective rate
\begin{gather} \label{eq:effratehighSNR-numerical}
\Ri = -\frac{1}{\theta TB_c}\log_e  \mathbb{E}\left\{ \exp\left(-\theta
T B_c \log_2 \det \left(\I + \frac{n_R}{n_T} \tsnr \hh \hh^\dd\right)\right)\right\}
\end{gather}
is plotted as a function of the bit energy. Above in (\ref{eq:effratehighSNR-numerical}), $B_c$ denotes the coherence bandwidth, and $\tsnr = \frac{P}{n_R m B_c N_0}$ where $m$ is the number of noninteracting subchannels, each experiencing i.i.d. zero-mean, unit-variance Gaussian fading. In this figure, we have $n_R = n_T = 2$, and $\frac{P}{N_0} = 10^4$, $T = 1$ms. We consider the setting in which the number of subchannels is bounded while the coherence bandwidth increases with increasing bandwidth. We assume $m = 5$ and plot the curves by varying $B_c$ from 10kHz to 10MHz. Therefore, bandwidth increases from 50kHz to 50MHz. As predicted by the result of Theorem \ref{theo:minbitenergywb}, the minimum bit energy depends on $\theta$ and increases with increasing $\theta$. We note that for relatively large values of $\theta$, considerably higher bit energies are needed when compared with the case of $\theta = 0$.

In Fig. \ref{fig:nR2nT2ebno-wideband}, we have assumed that the number of subchannels and hence the number of degrees of freedom is bounded, and $B_c$ increases linearly with increasing bandwidth. We have seen that having bounded number of degrees of freedom induces substantial energy penalty especially if the queueing constraints are stringent. Another scenario in sparse multipath fading is the one in which $B_c$ increases but only sublinearly with $B$. In such a case, the number of subchannels $m$ increases with $B$ as well. In Theorem \ref{theo:minbitenergywblargem}, we have shown for this scenario that the same minimum bit energy as in the case of $\theta = 0$ can be attained. This is depicted in Fig. \ref{fig:nR2nT2ebno-wideband_m}. In this figure, the parameters are the same as in Fig. \ref{fig:nR2nT2ebno-wideband}, except we now assume that $m$ increases from 5 to 100 as $B_c$ increases from 10kHz to 10MHz. We note that in all cases, the minimum bit energy of $-7.61$ dB is approached. However, it is interesting to observe that the wideband slopes are zero when $\theta > 0$, indicating that approaching the minimum bit energy is very demanding in terms of bandwidth in the presence of queueing constraints.

\section{Conclusion} \label{sec:conclusion}

In this paper, we have investigated the performance of MIMO wireless systems operating under statistical queueing (or QoS) constraints, which are formulated as limitations on buffer violation probabilities in the large-queue-length regime. We have employed effective capacity as the performance metric that provides the throughput under such constraints. We have studied the effective capacity in the low-power, wideband, and high-$\tsnr$ regimes. In the low-power regime, we have obtained expressions for the first and second derivatives of the effective capacity at zero $\tsnr$ under various assumptions on the channel knowledge at the transmitter side. We have shown that while the first derivative does not depend on the QoS constraints, the second derivative diminishes as these constraints become more stringent. As a byproduct of these results, we have demonstrated that the minimum bit energy requirements in the presence of QoS constraints in the low-power regime are the same as those required in the absence of such constraints. However, the wideband slope is shown to significantly get affected by queueing constraints.

Results derived in the low-power regime are proven to apply to the wideband regime in rich multipath fading environments. On the other hand, we have noted that sparse multipath fading induces energy penalty if the number of noninteracting subchannels remains bounded in the wideband regime. In this case, the minimum bit energy is shown to depend on the QoS exponent $\theta$. If the number of subchannels increase with bandwidth but only sublinearly, we have seen that the minimum bit energy required in the absence of buffer constraints can be attained, but we have demonstrated in the numerical results that approaching this level is very slow.

Finally, we have investigated the performance in the high-$\tsnr$ regime by determining the high-$\tsnr$ slope and power offset values. In particular, we have shown that if the QoS exponent is less than a certain thereshold, the high-$\tsnr$ slope of $\min(n_R,n_T)$ can be maintained. However, in this case, we have remarked that there is still a price to be paid in terms of the power offset $\li$ when queueing limitations are present. For the single-antenna case, we have proven that increasing $\theta$ beyond a threshold starts affecting the high-$\tsnr$ slope $\si$. In such a case, $\si$ is shown to diminish with increasing $\theta$. We have discussed extensions of this result to the multiple-antenna scenarios, and illustrated them through numerical results.

\end{spacing}

\end{document}